# Software Requirements Specification

## Unified University Inventory System (UUIS)

Prepared for : Imaginary University of Arctica

Version 1.0

Submitted to: Mr. Serguei Mokhov

By: Team1, COMP 5541, Winter 2010

Members of Team1:

| Name | Email |
|---|---|
| Abirami Sankaran | a_sankar@encs.concordia.ca |
| Andriy Samsonyuk | a_samso@encs.concordia.ca |
| Maab Attar | m_att@encs.concordia.ca |
| Mohammad Parham | m_parham@encs.concordia.ca |
| Olena Zayikina | o_zayik@encs.concordia.ca |
| Omar Jandali Rifai | o_jandal@encs.concordia.ca |
| Pavel Lepin | p_lepin@encs.concordia.ca |
| Rana Hassan | ra_hass@encs.concordia.ca |

# TABLE OF CONTENTS

## LIST OF TABLES

## LIST OF FIGURES

# 1. Introduction

This document, Software Requirements Specification (SRS), details the requirements to build a web based unified inventory system for the Imaginary University of Arctica (IUfA). The system, which facilitates the management of inventory for all the faculties of the University, is created to fulfill the requirements of COMP 5541 course Project. This document reports the requirements based on the official meetings of the members of Team 1 of COMP 5541, with the client Mr. Serguei Mokhov, the instructor of COMP 5541.

## 1.1 Purpose

This document provides the user characteristics and functions of the system, and specifies the details of all functional and quality requirements of the system. This document is intended for Mr. Serguei Mokhov and the designers, the web application developers, and the system testers of our team.

## 1.2 Scope

The web based inventory system, named Unified University Inventory System (UUIS), is unified for the use of all the faculties of the IUfA. The system requests the users, IUfA's students and staff, to log in to use the system. After authentication, the system allows the user to view the inventory of the IUfA, which is categorized as, Persons, Locations, Assets, and Licenses. The system allows the user to add, edit, and delete the items in the database (DB). The system also allows the user to request for one or more of the existing inventory or report on any such inventory.

The system restricts the users from accessing all the above mentioned features depending on the user's level of permissions. The level of permission is a default package, which is not assigned by the system. However, the system accepts change in the user's level or modifications made to the permissions in the present level of any user.

The system accepts data from the files of previous inventory system and the new data added by the user one after the other or by bulk operation.

## 1.3 Definitions, acronyms, and abbreviations

**Table 1.** Definitions

| Term/ Acronym/ Abbreviation | Expansion /Description |
|---|---|
| asset | Equipment (Computer, Keyboard, Monitor, Mouse, etc.), furniture (table, chair, grad seats, etc.), and software |
| auditing | Examination or verification of accounts |
| biometric characteristic | Analysis of biological data (voice) |
| bulk | More than one data |
| create | To make a new data in the database |
| group | Combine more than one entity |
| import | Add data to the system from a file |
| interface | An equipment or program for communication or interaction |
| inventory | All items (asset/license/person) that exists in the university |
| level | A hierarchy among the users |
| license | Legal instrument governing the usage or redistribution of software |
| location | Building, room, suites, cubicle, drawer, office, atriums, teaching labs, research labs |
| permission | The ability of the users to view the contents or make changes to the contents of the system |
| person | A person who has access to the system |
| profile | Collection of personal data associated to a specific user |
| role | User's designation like Student, Dean, Professor, IT administrator, etc. |
| system | The Unified University Inventory System |

**Table 2.** Acronyms/ Abbreviations

| | |
|---|---|
| CAPTCHA | Completely Automated Public Turing test to tell Computers and Humans Apart |
| CSV | Comma Separated Value |
| DB | Data base |
| DBMS | Database management system |
| GUI | Graphical User Interface |
| HTML | Hyper Text Markup Language |
| HTTP | Hypertext Transfer Protocol |
| IEEE | Institute of Electrical and Electronics Engineers |
| IUfA | Imaginary University of Arctica |
| JSP | Java Server Pages |
| JVM | Java Virtual Machine. |
| SRS | Software Requirements Specification |
| SSL | Syntax/Semantic Language. |
| UI | User interface. |
| UUIS | Unified University Inventory System |
| ZUI | Zooming user interface or zoomable user interface. |

## 1.5 References

IEEE Std 830-1998 IEEE Recommended Practice for Software Requirements Specifications, In *IEEE Xplore Digital Library*. http://ieeexplore.ieee.org/Xplore/guesthome.jsp

Mokhov, S. (2010). Selected Project Requirements. In *Concordia University*. http://users.encs.concordia.ca/~c55414/selected-project-requirements.txt

Team 1, 2010. Software Design Documentation, UUIS. Last Modified: Apr. 29, 2010.

Team 1, 2010. Software Test Documentation, UUIS. Last Modified: Apr. 29, 2010.

## 1.6 Overview

This document is written according to the standards for Software Requirements Specifications explained in "IEEE Recommended Practice for Software Requirements Specifications".
This document, in Section1.3, states the definition of certain words and expansion of abbreviations and acronyms used in the project, which could be used by all readers of the document to interpret the project correctly. The Specific requirements in Section 3, gives the details of the expected functions in the UUIS, intended for the developers in building phase and the test team, in testing and validating the final product. The appendix contains the estimated cost in developing this project.

# 2. Overall description

## 2.1 Product perspective

UUIS is an independent product that does not require additional hardware or software interfaces to function, other than the OS. When released, the final product would be the first version of the software. It is designed as a secured system, which could be accessed by the any authenticated user. Nonetheless, the system restricts access to its various components, to users with varied characteristics.

### 2.1.1. System interfaces

The data from previous inventory system are available in CSV format. UUIS will need an API to import data from those files.

### 2.1.2. User interfaces

The UI should be easy to manipulate without additional training. The user should be able to interact with the system in any of the languages available in the language menu. The pages should use a ZUI graphical environment, should be built with a good sense of color and contrast, and should be printable, using keys. All pages of the system should be accessible from any page.

### 2.1.3 Communication interfaces

The system requires HTTP to communicate with server. The system can be configured to be accessed via any available port.
The web based UI is the only means of communication between the user and the system. The system is accessible through all popular web browsers that interact with JSP and HTML pages.

### 2.1.4 Memory constraints

UUIS requires a minimum of 512 MB of primary memory and 3GB of secondary memory for installation and execution.

## 2.2. Product functions

The system performs the following functions. The functions depend on the user's level and permission package, as explained in the user characteristics.

### 2.2.1. Login

This function allows the user to enter into the application. The user is required to provide username and password. High privileged user in addition is required to provide a biometric characteristic. After authentication user will have access to main menu. Availability of menu functions depends on user's level and permission package.

### 2.2.2. View/add/edit/delete records

This function allows the user with appropriate permissions to view/add/edit/delete records in the appropriate categories of the DB. System has four main databases: “Persons”, “Assets”, “Locations”, and “Licenses”. The user could also make “Bulk entry”. Deleted record just changes status.

### 2.2.3. Create a group of assets/locations

This function allows the user with appropriate permissions to combine list of assets/locations into a group.

### 2.2.4. Create new subgroup of assets

This function allows the user with appropriate permissions to create a new subgroup of assets.

### 2.2.5. Create new type of asset/license/location/person

This function allows the user with appropriate permissions to create a new type of asset/license/location/person.

### 2.2.6. Import of assets/licenses/locations/persons (accounts)

This function allows the user with appropriate permissions to transfer data into DB from other systems. Import is possible to all main DB: “Persons”, “Assets”, “Locations”, and “Licenses”. User maps the fields manually. If there is no corresponding field, system will create “problem” file, which the user will check manually.

### 2.2.7. Add new role (package of permissions)

This function allows the administrator to create default package of permissions for users. There are different packages in the system.

### 2.2.8. Edit role

This function allows the administrator to change default package of permissions for the user who requested.

### 2.2.9. Add new permission

This function allows the administrator to add new permission, in case of modification to system’s functionality.

### 2.2.10. Edit permission

This function allows the administrator to correct the name of permission, before the permission is used in the system.

### 2.2.11. Add new request

This function can be done by all users. The users can report a problem, report a bug, make general requests, and move item(s). After submission, the requests will enqueue to a pool for approval/rejection. If a person in level 3 submits request, request is approved automatically.

### 2.2.12. Borrow asset(s)/license(s)

This function allows the user with appropriate permissions to borrow asset(s)/license(s). Basically this is assigning asset(s)/license(s) to person.

### 2.2.13. View list of all requests in the system

This function allows the user with appropriate permission to browse list of all requests existing in the system.

### 2.2.14. Approve/reject request

This function allows the user with appropriate permissions to approve/reject any submitted request. Every request needs approval by one level above. Person who rejects request shall provide reason for rejection.

### 2.2.15. Assign asset/license/location/person TO asset/license/location/person/role

This function allows the user with appropriate permissions to assign:

- asset(s) to person (bulk operation is available)
- asset(s) to location (bulk operation is available)
- license to asset
- location(s) to location (bulk operation is available)
- permission to person(s) (bulk operation is available)
- role to person(s) (bulk operation is available)

### 2.2.16. Search

This function enables the user to perform two kinds of search: basic search and advanced search. In basic search, the user types a query in a textbox and clicks on “Search” button. In advanced search, the user can constrain the search by one or more attributes that are available to choose from. In addition, the user can also make combinations of the chosen attributes by choosing AND, OR, or NOT. In both cases, the entry is searched as a case insensitive substring.

### 2.2.17. View/print plan of big location

There is a list of locations, which has plan (map). This function allows the user with appropriate permissions to select location in order to view plan (map). When user moves the mouse, could see the number and type of room and to whom it is assigned. HTMP mapping updates dynamically (if the room is reassigned to another person, the changes are seen in the map too).

### 2.2.18. View “My profile”

This function allows all the users to view list of all assets/license/locations assigned to him/her, and list of assets/license borrowed by him/her.

### 2.2.19. Provide biometrical characteristic

This function allows a high privileged user to provide biometrical characteristic (sample of voice) which will be used to login into the system.

### 2.2.20. Create/print reports

There is a standard list of reports in the system. This function allows the user with appropriate permission to create/print report. For type of location: teaching lab, research lab, office there are reports:

- compare capacity of location with number of chairs;
- compare capacity of location with number of tables;
- compare capacity of location with number of PC (only for teaching labs and offices);
- compare capacity of location with number of students (only for research lab);

### 2.2.21. Help

This function guides the user on how to use a particular page or functions in that page.

### 2.2.22. Auditing

Every action in the system is saved in a database. A user with auditing permission is allowed to access the database. The user could view all the records or view filtered records by specifying a period of time or selecting asset/location/license/person.

### 2.2.23. Logout

This function can be done by all users. It terminates the user session. The system can also do this function automatically if the session is left unused for half an hour.

## 2.3. User characteristics

The users of this application are assigned one of the following levels:
The University level, the faculty level, the department level and users level. We call them level 3, level 2, level 1 and level 0 respectively.
Leveling is mainly used as area of visibility in order to show data for University, Faculty, and Department. As an example, a level admin is able to manage departmental assets inside the department while a faculty admin is not only able to manage departmental assets but also manage faculty level assets among all departments of the faculty.
In addition, leveling is used for approval/rejection of requests. Once a user of some level (n) submits a request, it is generally seen, approved, or rejected by a user/admin in level (n+1). However, if there exists a level above (n+1), a user/admin of that level could also see, approve or reject a request. Exceptionally, requests of level 3 users need no approval from anyone. The requests are approved automatically on submission.
Same role in different levels has different permissions. For example level 2 of admin (faculty admin) is allowed to manage faculty inventory, searching at the faculty level for anything in the inventory (for example if he wants to move a printer at the faculty level) and obviously has access to all the rooms of faculty building. But level 1 of admin has all these permissions from another scope (one level less, at the departmental level).
As from a general top view, users are grouped in one of the following groups:

- University employees
- IT administrators
- Faculty members
- Students
- Visitors

Each group may have subgroups with different levels of authority. Each subgroup (role) in turn has different permissions according to the general application description. Although the set of permissions for each role could be changed in the system, typical set of permissions for each role is defined in the table below.

**Table 3.** User characteristics

| Role | # | Permission |
|---|---|---|
| **administrator** | 1 | insertAsset |
| | 2 | seeAssets |
| | 3 | editAsset |
| | 4 | deleteAssets |
| | 5 | borrowAssets |
| | 6 | addGroupAsset |
| | 7 | addTypeAsset |
| | 8 | addSubgroupAsset |
| | 9 | importAsset |
| | 10 | assignAssetsToPerson |
| | 11 | assignAssetsToLocation |
| | 12 | seeMyAssets |
| | 13 | insertLocation |
| | 14 | seeLocations |
| | 15 | editLocation |
| | 16 | deleteLocations |
| | 17 | addGroupLocation |
| | 18 | addTypeLocation |
| | 19 | see_printFloorPlan |
| | 20 | importLocation |
| | 21 | assignLocationToPerson |
| | 22 | assignLocationToLocation |
| | 23 | assignLocationToDepartment |
| | 24 | seeMyLocations |
| | 25 | insertLicense |
| | 26 | seeLicenses |
| | 27 | editLisense |
| | 28 | deleteLicenses |
| | 29 | borrowLicenses |
| | 30 | addTypeLicence |
| | 31 | importLicense |
| | 32 | assignLicenceToAsset |
| | 33 | seeMyLicenses |
| | 34 | seePersons |
| | 35 | editPerson |
| | 36 | deletePersons |
| | 37 | addBiometric |
| | 38 | importPerson |
| | 39 | addRole |
| | 40 | editRole |
| | 41 | addPermission |
| | 42 | editPermission |
| | 43 | assignPermissionToPersons |
| | 44 | assignRoleToPersons |
| | 45 | seeMyRole |
| | 46 | seeMyPermissions |
| | 47 | insertFacDep |
| | 48 | seeFacDep |
| | 49 | editFacDep |
| | 50 | createAcquisitionRequest |
| | 51 | createReparationRequest |
| | 52 | createEliminationRequest |
| | 53 | createMoveRequest |
| | 54 | aprove_rejectRequest |
| | 55 | seeRequestsAll |
| | 56 | basicSearch |
| | 57 | advancedSearch |
| | 58 | create_printReport |
| | 59 | seeAudit |
| | 60 | seeMyProfile |
| | 61 | selectLanguage |
| | 62 | login_logout |
| **similar for all roles (not repeated for next roles)** | 1 | basicSearch |
| | 2 | createAcquisitionRequest |
| | 3 | createEliminationRequest |
| | 4 | createReparationRequest |
| | 5 | login_logout |
| | 6 | seeMyAssets |
| | 7 | seeMyLicenses |
| | 8 | seeMyLocations |
| | 9 | seeMyProfile |
| | 10 | seeMyPermissions |
| | 11 | seeMyRole |
| | 12 | selectLanguage |
| **full_time_faculty** | 1 | addBiometric |
| | 2 | borrowAssets |
| | 3 | borrowLicenses |
| | 4 | create_printReport |
| | 5 | createMoveRequest |
| | 6 | deleteAssets |
| | 7 | deleteLicenses |
| | 8 | deleteLocations |
| | 9 | deletePersons |
| | 10 | editAsset |
| | 11 | editFacDep |
| | 12 | editLisense |
| | 13 | editLocation |
| | 14 | editPerson |
| | 15 | importAsset |
| | 16 | importLicense |
| | 17 | importLocation |
| | 18 | importPerson |
| | 19 | insertAsset |
| | 20 | insertFacDep |
| | 21 | insertLicense |
| | 22 | insertLocation |
| | 23 | see_printFloorPlan |
| | 24 | seeAssets |
| | 25 | seeAudit |
| | 26 | seeFacDep |
| | 27 | seeLicenses |
| | 28 | seeLocations |
| | 29 | seePersons |
| | 30 | seeRequestsAll |
| **part_time_faculty** | 1 | addBiometric |
| | 2 | advancedSearch |
| | 3 | borrowAssets |
| | 4 | borrowLicenses |
| | 5 | create_printMoveRequest |
| | 6 | create_printReport |
| | 7 | create_printReport |
| | 8 | see_printFloorPlan |
| | 9 | seeAssets |
| | 10 | seeLicenses |
| | 11 | seeLocations |
| | 12 | seePersons |
| **full_time_worker** | 1 | advancedSearch |
| | 2 | see_printFloorPlan |
| | 3 | seeAssets |
| | 4 | seeLicenses |
| | 5 | seeLocations |
| | 6 | seePersons |
| **part_time_worker** | 1 | see_printFloorPlan |
| | 2 | seeAssets |
| | 3 | seeLicenses |
| | 4 | seeLocations |
| **grad_student** | 1 | advancedSearch |
| | 2 | borrowAssets |
| | 3 | borrowLicenses |
| | 4 | createMoveRequest |
| | 5 | see_printFloorPlan |
| | 6 | seeAssets |
| | 7 | seeLicenses |
| | 8 | seeLocations |
| | 9 | seePersons |
| **undergrad_student** | 1 | see_printFloorPlan |
| **independent_student** | 1 | see list of permissions similar for all roles |

## 2.4. Constraints

UUIS is implemented in Java. To install and execute the inventory system, JVM is required. Additionally, the system requires Tomcat and phpMyAdmin. The system supports up to 200 users at the same time.

## 2.5. Assumptions and dependencies

The personal information - ID, title, username (login), communication information, user's role, and user's level in the system - is previously determined for all the users of the university. Access to the system is restricted only to the pre-assigned usernames (logins). Higher level users are aware of the restrictions and limitations entitled to all the levels. Any security personnel can access/verify any user's account in the university. Every action taking place in the system is required to be saved in a separate database, for auditing purposes.

## 2.6. Apportioning of requirements

V.1 of the system is available only in English.  While functions in Faculty and Department await future versions of the system, major functions in other categories are available. The SDD of UUIS can be referred for a detailed list of available functions.

# 3. Specific requirements

## 3.1. External interface requirements

The system takes input from scanner, keyboard, and files in the memory. The system generates printable output on the screen and peripherals. The system uses MySQL database and JDBC connector to communicate with database.

## 3.2. Functional requirements

### 3.2.1. Use cases

The use cases describe the procedures and exemptions for each function. The appropriate permissions of each user are listed in section 2.3. (User Characteristics) of this document.

### 3.2.1.1 Add new asset

**Actor(s)**: Users with appropriate permissions.
**Precondition(s)**:

1. Authenticated session.
2. Appropriate request is submitted and approved.
3. No such asset in the DB.

**Trigger**: Clicking menu “Asset ➔ Add new”.

**Procedure:**

1. User on the page “Main” clicks on menu “Asset ➔ Add new” and system transfers in to page “Asset ➔ Add new”.
2. System opens DB “Asset”.
3. User selects type of assets (from available list) and system automatically shows appropriate characteristics.
4. System automatically generates created date.
5. User provides following information: name, subgroup (from list if available), serial number, barcode, purchase number, request number, color (for furniture), material (for furniture), host name (for computers), status (from available list), brand, version (for software) and description.
6. User selects (assigns) location from available list.
7. User terminates operation:

    7.1. User clicks the button “Create” to save changes.

        7.1.1. System checks if barcode is unique.

        7.1.2. System updates the action in Audit file.

    7.2. User clicks the button “Cancel” to abort changes.

**Post condition(s):**

1. New asset is added to DB “Asset”.

**Exception(s):**

2a. Database error

2a1. System displays an error message

7.1a. Required field(s) is (are) empty.

7.1a1. System displays an error message and ask user to repeat again.

7.2a. Barcode is not unique.

7.2a1. System displays an error message and asks user to change barcode.

### 3.2.1.2. View DB “Asset” / DB “License” / DB “Location” / DB “Person” / Faculty / Department

**Actor(s)**: Users with appropriate permission.

**Precondition(s)**:

1. Authenticated session.
2. Appropriate nonempty DB exists in the UUIS.

**Trigger**: Clicking menu “Asset ➔ View”/ “License ➔ View”/ “Location ➔ View”/ “Person (account) ➔ View”/ “Faculty ➔ View”/”Department➔View”.

**Procedure:**

1. User on page “Main”:

 1.1. User on the page “Main” clicks on menu “Asset ➔ View” and system transfers to page “Asset ➔ View”.

 1.2. User on the page “Main” clicks on menu “License ➔ View” and system transfers to page “License ➔ View”.

 1.3. User on the page “Main” clicks on menu “Location ➔ View” and system transfers to page “Location ➔ View”.

 1.4. User on the page “Main” clicks on menu “Person ➔ View” and system transfers to page “Person ➔ View.

 1.5. User on the page “Main” clicks on menu “Person ➔ View” and system transfers to page “Faculty ➔ View”.

1.6. User on the page "Main" clicks on menu "Person ➔ View" and system transfers to page "Department ➔ View".

2. System displays:

2.1. System displays DB "Asset" in a table. Displayed column depends on role and level of user. Displayed lines depend on Faculty and Department to which user belongs.

2.2. System displays DB "License" in a table. Displayed column depends on role and level of user. Displayed lines depend on Faculty and Department to which user belongs.

2.3. System displays DB "Location" in a table. Displayed column depends on role and level of user. Displayed lines depend on Faculty and Department to which user belongs.

2.4. System displays DB "Person" in a table. Displayed column depends on role and level of user. Displayed lines depend on Faculty and Department to which user belongs.

2.5. System displays list of Faculties in a table. Displayed column depends on role and level of user. Displayed lines depend on Faculty to which user belongs.

2.6. System displays list of Departments in a table. Displayed column depends on role and level of user. Displayed lines depend on Faculty and Department.

3. The user clicks on a record to see the details. The system displays the details of the record in a new page.

4. System updates the action in Audit file.

**Post condition(s):**

1. Records in appropriate DB are viewed.

**Exception(s):**

2a. Database error.

2a1. System displays an error message.

### 3.2.1.3. Edit record(s) in DB "Asset" / DB "License" / DB "Location" / DB "Person" / Faculty / Department

**Actor(s)**: Users with appropriate permission.

**Precondition(s)**:

1. Authenticated session.

2. Appropriate nonempty DB exists in the UUIS.

3. Appropriate request is submitted and approved.

**Trigger**: Clicking link “Asset → Edit”/ “License → Edit”/ “Location → Edit”/ “Person (account) → Edit”.

**Procedure:**

1. User on the page “Main”:
    1.1. User on the page “Main” clicks on menu “Asset” and system transfers to page “Asset”.
    1.2. User on the page “Main” clicks on menu “License” and system transfers to page “License”.
    1.3. User on the page “Main” clicks on menu “Location” and system transfers to page “Location”.
    1.4. User on the page “Main” clicks on menu “Person” and system transfers to page “Person”.
2. System displays:
    2.1. System displays DB “Asset” in the table. User can show/hide columns, change number of lines per page. Display depends on role, level, Faculty, and Department to which user belongs.
    2.2. System displays DB “License” in the table. User can show/hide columns, change number of lines per page. Display depends on role, level Faculty, and Department to which user belongs.
    2.3. System displays DB “Location” in the table. User can show/hide columns, change number of lines per page. Display depends on role, level Faculty, and Department to which user belongs.
    2.4. System displays DB “Person” in the table. User can show/hide columns, change number of lines per page. Display depends on role, level Faculty, and Department to which user belongs.
3. User clicks on “Edit” link next to the record which should be modified.
4. System opens new window which contains only one record.
5. User starts editing.
6. User clicks on “Submit” button to save changes or button “←” to abort changes.
    6.1. System checks if barcode is unique.

6.2. System updates the action in Audit file.

**Post condition(s):**

1. Records in appropriate DB are modified.

**Exception(s):**

2a. Database error.

2a1. System displays an error message.

6a. Compulsory field(s) is (are) empty.

6a1. System displays an error message and ask user to repeat again.

6.1a. Barcode is not unique.

6a1. System displays an error message and asks user to change barcode.

#### 3.2.1.4. Delete record(s) in DB “Asset” / DB “License” / DB “Location” / DB “Person”

**Actor(s)**: Users with appropriate permission.

**Precondition(s)**:

1. Authenticated session.
2. Appropriate nonempty DB exists in the UUIS.

**Trigger**: Clicking menu “Asset ➔ Delete”/ “License ➔ Delete”/ “Location ➔ Delete”/ “Person (account) ➔ Delete”.

**Procedure:**

1. User on the page “Main”:

   1.1. User on the page “Main” clicks on menu “Asset ➔ Delete” and system transfers to page “Asset ➔ Delete”.

   1.2. User on the page “Main” clicks on menu “License ➔ Delete” and system transfers to page “License ➔ Delete”.

   1.3. User on the page “Main” clicks on menu “Location ➔ Delete” and system transfers to page “Location ➔ Delete”.

1.4. User on the page “Main” clicks on menu “Person ➔ Delete” and system transfers to page “Person ➔ Delete”.

2. System displays:

   6.3. System displays DB “Asset” in the table. User can show/hide columns, change number of lines per page. Displayed information depends on role, level, Faculty, and Department to which user belongs.

   6.4. System displays DB “License” in the table. User can show/hide columns, change number of lines per page. Displayed information depends on role, level, Faculty, and Department to which user belongs.

   6.5. System displays DB “Location” in the table. User can show/hide columns, change number of lines per page. Displayed information depends on role, level, Faculty, and Department to which user belongs.

   6.6. System displays DB “Person” in the table. User can show/hide columns, change number of lines per page. Displayed information depends on role, level, Faculty, and Department to which user belongs.

3. User uses scroll bar in order to see different records in the table.
4. User selects record(s) which should be deleted by clicking on appropriate checkbox (es).
5. User clicks on “Delete Selected” button.
6. System shows the message “Please select assets to delete and press “Delete Selected” button”.
7. User clicks on the checkbox (es) next to the records that should be deleted.
8. User presses “Delete Selected” button.
9. System shows message that record(s) was/were deleted (physically the record(s) still present in DB, only the status is changed to “unavailable”).
10. System updates the action in Audit file.

**Post condition(s):**

1. Record(s) in appropriate DB are deleted.

**Exception(s):**

2a. Database error.

2a1. System displays an error message.

8a. User didn't select any records in DB "Asset".

8a1. System displays an error message and ask user to select records in DB "Asset".

### 3.2.1.5. Create a group of assets / locations

**Actor(s)**: Users with appropriate permission.

**Precondition(s)**:

1. Authenticated session.
2. There is no such group of assets/locations in DB "Asset"/DB "Location".

**Trigger**: Clicking menu "Asset ➔ Create a group"/ "Location ➔ Create a group".

**Procedure:**

1. User on the page "Main" clicks on menu "Asset ➔ Create a group"/ "Location ➔ Create a group" and system transfers in to page "Asset ➔ Create a group"/ "Location ➔ Create a group".
2. System displays DB "Asset"/ DB "Location" in the table. User can show/hide columns, change number of lines per page. Displayed information depends on role, level, Faculty, and Department to which the user belongs.
3. User uses scroll bar in order to see different records in the table.
4. System asks user to select "Asset master" (name of group)/ "Location master", clicking on appropriate checkbox.
5. User selects "Asset master" /"Location master".
6. User selects next step.

   6.1. User clicks on button "Add master", in order to confirm operation.

   6.2. User clicks on button "Cancel", in order to cancel operation. In this case user terminates process.
7. System asks user to select "Asset children" (component(s) of group)/ "Location children", clicking on appropriate checkbox (es).
8. User selects next step.

8.1. User clicks on button “Add Children”, in order to confirm operation. User has opportunity to repeat this step many times as necessary.

8.2. User clicks on button “Cancel”, in order to cancel operation. In this case user terminates process.

9. User terminates operation:

9.1. User clicks “Create a group” button in order to confirm creating group of assets/locations.

9.1.1. System updates the action in Audit file.

9.2. User can click “Cancel” button to abort creating new group of assets/locations.

**Post condition(s):**

1. New group of assets/locations is created in the DB “Asset”/ DB “Location”.

**Exception(s):**

2a. Database error

2a1. System displays an error message

6.1a. User selects more than one “Asset master”/ “Location master”.

6.1a1. System displays an error message and asks user to select only one “Asset master”/ “Location master”.

8.1a. User selects only one “Asset children”/ “Location children” which should be included into group.

8.1a1. System displays an error message and asks user to select more than one asset/location which should included into group.

### 3.2.1.6. Create new type of asset / license / location / person

**Actor(s)**: Users with appropriate permission.

**Precondition(s)**:

1. Authenticated session.
2. There is no such type of asset/license/location/person.

**Trigger**: Clicking menu “Asset ➔ Create new type”/ “License ➔ Create new type”/ “Location ➔ Create new type”/ “Person ➔ Create new type”.

**Procedure:**

1. User on the page “Main”:

   1.1 User on the page “Main” clicks on menu “Asset ➔ Create new type” and system transfers to page “Asset ➔ Create new type”/

   1.2 User on the page “Main” clicks on menu “License ➔ Create new type” and system transfers to page “License ➔ Create new type”/

   1.3 User on the page “Main” clicks on menu “Location ➔ Create new type” and system transfers to page “Location ➔ Create new type”/

   1.4 User on the page “Main” clicks on menu “Location ➔ Create new type” and system transfers to page “Person ➔ Create new type”.

2. User types name of the new type of asset/license/location/person in the appropriate field.

3. User selects fields which should be included in the interface for particular type of asset/license/location/person.

4. User terminates operation:

   4.1. User clicks “Submit” button to save the name of new type.

      4.1.1. System automatically adjusts the list of available types of asset/license/location in the DB “Asset”/ DB “License”/ DB “Location”/DB “Person”.

      4.1.2. System updates the action in Audit file.

   4.2. User clicks on “Cancel” to abort the operation.

**Post condition(s):**

1. New type of asset/license/location/person is created.

**Exception(s):**

2a. User typed the name of type, which already exists.

2a1. System displays an error message and ask user to rewrite name of type.

4.1a. User didn’t write name of type.

4.1a1. System displays an error message and asks user to write name of type.

4.1b. User didn’t select compulsory fields which should be included in interface for particular type of asset/license/location/person.

4.1b1. System displays an error message and asks user to select compulsory fields.

### 3.2.1.7. Create new subgroup of assets

**Actor(s)**: Users with appropriate permission.

**Precondition(s)**:

1. Authenticated session.
2. There is no such subgroup of asset.

**Trigger**: Clicking menu “Asset ➔ Create new subgroup”.

**Procedure:**

1. User on the page “Main” clicks on menu “Asset ➔ Create new subgroup” and system transfers to page “Asset ➔ Create new subgroup”.
2. System displays DB “Asset” in a table. User can show/hide columns, change number of lines per page. Displayed information depends on role, level, Faculty, and Department to which user belongs.
3. User selects record(s) which should be included into subgroup, putting tick(s) in appropriate checkbox (es).
4. User writes the name on subgroup in appropriate field.
5. User terminates operation:

    5.1. User clicks on “Submit” button to confirm creation of new subgroup.

    5.1.1. System automatically adjusts list of available subgroups of assets in the DB “Assets”.

    5.1.2. System automatically updates the Audit file.

    5.2. User clicks on “Cancel” to abort operations.

**Post condition(s):**

1. New subgroup of assets is created.

**Exception(s):**

2a. Database error.

2a1. System displays an error message.

4a. User wrote the name of subgroup, which already exists.

4a1. System displays an error message and asks user to rewrite name of type.

5.1 User didn't select record(s) which should be included into subgroup.

5.1a System displays an error message and asks user to select record(s) which should be included in the subgroup.

4.1a User didn't write name of subgroup.

4.1a1 System displays an error message and asks user to write name of subgroup.

### 3.2.1.8. Import of assets / licenses / locations / persons (accounts)

**Actor(s)**: Users with appropriate permission.

**Precondition(s)**:

1. Authenticated session.
2. Necessity to integrate old data (asset/license/location/person) with UUIS.

**Trigger**: Clicking menu "Asset ➔ Import"/ "License ➔ Import"/"Location ➔ Import"/ "Person ➔ Import".

**Procedure:**

1. User on the page "Main":

   1.1. User on the page "Main" clicks on menu "Asset ➔ Import" and system transfers to page "Asset ➔ Import"/

   1.2. User on the page "Main" clicks on menu "License ➔ Import" and system transfers to page "License ➔ Import"/

   1.3. User on the page "Main" clicks on menu "Location ➔ Import" and system transfers to page "Location ➔ Import"/

1.4. User on the page “Main” clicks on menu “Person ➔ Import” and system transfers to page “Person ➔ Import”.

2. User selects type of import:

2.1. From file (for assets, licenses, locations, persons)

2.1.1. User opens *.csv or *.txt file for import.

2.1.2. User makes a copy the contents of *.csv or *.txt file and pastes in appropriate textbox.

2.2. From scanner or keyboard (for assets and license)

2.2.1. User scans items.

2.2.2. User manually provide style of *.csv or *.txt file.

3. User, from the available list, manually selects the corresponding name of the field(s) and puts number of corresponding column.

4. In drop down menu user selects location (if it is export for assets or licenses) where items should be imported.

5. User terminates operation:

5.1. User clicks “Upload inserted data” button to add file’s contents to the appropriate DB.

6.1.1. System inserts assets/licenses/locations/persons into appropriate DB.

6.1.2. System automatically generates created date

6.1.3. System automatically updates the Audit file.

5.2. User clicks on “Cancel” button to abort import of assets/licenses/locations/persons.

6. User manually fills empty fields in inserted records in appropriate database.

**Post condition(s):**

1. Assets/licenses/locations/persons from file are imported into UUIS.

**Exception(s):**

5.1a. Data in textbox is in incorrect format.

5.1a1. System displays an error message and asks user to select correct format.

5.1b. User didn’t select location to perform “import”.

5.1b1. System displays an error message and asks user to select location.

5.1c. Database error.

5.1c1. System displays an error message.

### 3.2.1.9. Assign asset(s) to person / location (bulk operation included)

**Actor(s)**: Users with appropriate permission.

**Precondition(s)**:

1. Authenticated session.
2. Appropriate request is submitted and approved.

**Trigger**: Clicking menu “Asset ➔ Assign to person/location”.

**Procedure:**

1. User on the page “Main” clicks on menu “Asset ➔ Assign to person/location” and system transfers in to page “Asset ➔Assign to person/location”.
2. System displays DB “Asset” in a table. User can show/hide columns, change number of lines per page. Displayed information depends on role, level, Faculty, and Department to which the user belongs.
3. In appropriate list user selects location/writes username of person to which asset(s) should be assign. Displayed list of locations depends on role, level, Faculty, and Department to which the user belongs.
4. User selects:

4.1. User manually selects asset(s) which should be assigned to person/location.

4.1.1. User clicks on appropriate checkbox (es).

4.1.2. User uses scroll bar in order to see different records in the table.

4.2. User scans asset(s) which should be assigned to person/location.

4.2.1. User clicks “Load/scan items”

4.2.2. User scans barcodes of assets to appropriate textbox.

4.2.3. System automatically reads barcodes into appropriate textbox.

4.3. User uses *.csv/*.txt file with list of asset(s) which should be assigned to person/location.

4.3.1. User clicks “Load/scan items”

4.3.2. User opens the *.csv/*.txt file.

4.3.3. User copies all barcodes from the *.csv/*.txt file to appropriate textbox.

5. User assigns:

5.1. User clicks “Assign to location” / “Assign to user” / “Submit” button to assign asset(s) to person/location.

5.1.1. System verifies if asset belong to Faculty/Department to which the user belongs.

5.1.2. System updates fields “Assign to person/Assign to location” and “Date” in DB “Asset”.

5.1.3. System automatically updates the audit file.

5.2. User can click on “Cancel” to abort assigning asset(s) to person/location.

**Post condition(s):**

1. Assets assigned to person/ location.

**Exception(s):**

2a. Database error.

2a1. System displays an error message.

3a. User didn’t select person/location to which selected assets should be assigned.

3a1. System displays an error message and asks user to select person/location to which selected assets should be assigned.

4.1.1a. User didn’t select any records in DB “Asset”.

4.1.1a1. System displays an error message and asks user to select records in DB “Asset”.

5.1a. User selects asset(s), which already assigned to another person/location.

5.1a1. System displays an error message and asks user to select another asset(s) in DB “Asset”.

5.1.1a. Scanned asset or asset from file doesn’t belong to the same Faculty/Department as user. (User doesn’t have right to move this asset).

5.1.1a1. System displays an appropriate message and tells that user doesn't have right to move this asset.

### 3.2.1.10. Borrow asset(s) / license(s)

**Actor(s)**: Users with appropriate permission.

**Precondition(s)**:

1. Authenticated session.
2. Appropriate request is submitted and approved.

**Trigger**: Clicking menu "Asset ➔ Borrow"/ "License ➔ Borrow".

**Procedure:**

1. User on the page "Main" clicks on menu "Asset ➔ Borrow"/ "License ➔ Borrow" and system transfers in to page "Asset ➔ Borrow" / "License ➔ Borrow".
2. System displays DB "Asset"/"License" in a table. User can show/hide columns, change number of lines per page. Displayed information depends on role, level, Faculty, and Department to which the user belongs.
3. In appropriate list, user selects location/writes username of person to which asset(s) should be assign. Displayed list of locations depends on role, level, Faculty, and Department to which the user belongs.
4. User selects asset(s)/license(s) (by clicking on appropriate checkbox(es)) the requester wants to borrow.
5. User terminates operation:

5.1. User clicks "Submit" button in order to confirm borrowing asset(s)/license(s).

5.1.1. System updates status of asset(s)/license(s).

5.1.2. System automatically sends e-mail to requester, asking him/her to pick up requested item(s).

5.1.3. System automatically updates the audit file.

5.2. User clicks on "Cancel" to abort loan.

**Post condition(s):**

1. Asset(s)/license(s) assigned to person.

2. Status of asset(s)/license(s) changed.

**Exception(s):**

2a. Database error.

2a1. System displays an error message.

5.1a. User didn’t select person which wants to borrow asset(s)/license(s).

5.1a1. System displays an error message and asks user to select person that wants to borrow asset(s)/license(s).

5.1b. User didn’t select any records in DB “Asset”/”License”.

5.1b1. System displays an error message and asks user to select records in DB “Asset”/”License”.

## 3.2.1.11. Add new license

**Actor(s)**: Users with appropriate permission.

**Precondition(s)**:

1. Authenticated session.
2. No such license in the DB “License”.
3. Appropriate request is submitted and approved.

**Trigger**: Clicking menu “License ➔ Add new”.

**Procedure:**

1. User on the page “Main” clicks on menu “License ➔ Add new” and system transfers to page “License ➔ Add new”.
2. System automatically generates created date.
3. User provides following information: name, purchase number, request number, type (from available list), number of available license, price, term, name of company.
4. User terminates operation:

4.1. User clicks the “Create” button to save changes.

4.1.1. System automatically updates the audit file.

4.2. User aborts changes by clicking on “Cancel” button.

**Post condition(s):**

1. New license added into DB “License”.

**Exception(s):**

1a. Database error

1a1. System displays an error message

4.1a. Compulsory field(s) is (are) empty.

4.1a1. System displays an error message and asks user to repeat again.

## 3.2.1.12. Assign license to asset

**Actor(s)**: Users with appropriate permission.

**Precondition(s)**:

1. Authenticated session.
2. Appropriate request is submitted and approved.

**Trigger**: Clicking link “Assign it”.

**Procedure:**

1. User on the page “Main” clicks on menu “License ➔ Assign to asset” and system transfers to page “License ➔ Assign to asset”.
2. System displays DB “License” in a table. User can show/hide columns, change number of lines per page. Displayed information depends on role, level, Faculty, and Department to which the user belongs.
3. In DB user selects license to which asset should be assigned, by clicking link “Assign it” next to the appropriate license. Displayed list of license depends on role, level, Faculty, and Department to which the user belongs.
4. In appropriate list user selects asset to which license should be assigned.
5. User terminates operation:

5.1. User clicks on “Assign” button to assign license to asset.

5.1.1. System automatically updates the audit file.

5.2. User clicks on “🡐” button to abort assigning license to asset.

**Post condition(s):**

1. License assigned to asset.

**Exception(s):**

2a. Database error.

2a1. System displays an error message.

5a. User didn’t select license which should be assigned to asset.

5a1. System displays an error message and asks user to select license which should be assigned to asset.

5b. User didn’t select asset to which license should be assigned.

5b1. System displays an error message and asks user to select asset to which license should be assigned.

### 3.2.1.13. Add new location

**Actor(s)**: Users with appropriate permission.

**Precondition(s)**:

1. Authenticated session.
2. No such location in the DB “Location”.
3. Appropriate request is submitted and approved.

**Trigger**: Clicking menu “Location ➔ Add new”.

**Procedure:**

1. User on the page “Main” clicks on menu “Location ➔ Add new” and system transfers to page “Location ➔ Add new”.
2. System automatically generates created date.
3. User provides following information: type (from available list), capacity, description, location number, key number, code number, width, length, belong to: University, Faculty (from available list)/ Department (from available list).
4. User terminates operation:

   4.1. User saves changes by clicking on “Create” button.

4.1.a. System automatically updates the audit file.

4.2. User aborts changes by clicking on the “Cancel” button.

**Post condition(s):**

1. New location added into DB “Location”.

**Exception(s):**

1a. Database error

1a1. System displays an error message

3a. Compulsory field(s) is (are) empty.

3a1. System displays an error message and asks user to repeat again.

### 3.2.1.14. Assign location(s) to another location /department (bulk operation included)

**Actor(s)**: Users with appropriate permission.

**Precondition(s)**:

1. Authenticated session.
2. Appropriate request is submitted and approved.

**Trigger**: Clicking menu “Location ➔ Assign to location”.

**Procedure:**

1. User on the page “Main” clicks on menu “Location ➔ Assign to location” and system transfers to page “Location ➔ Assign to location”.

6. System displays DB “Location” in the table. Displayed information depends on role, level, Faculty, and Department to which the user belongs.

2. User uses scroll bar in order to see different records in the table.

3. User selects location(s) which should be assigned to another location/Department by clicking on appropriate checkbox (es).

4. User selects (from available list) another location/Department to which selected location(s) should be assigned.

5. User terminates operation:

5.1. User clicks on “Assign” button in order to assign location(s) to person/location.

5.1.1. System updates fields “Assign to location”/”Assign to Department” and “Date” in DB “Location”.

5.1.2. System automatically updates the audit file.

5.2. User clicks on “Cancel” button to abort assigning location(s) to person/location.

**Post condition(s):**

1. Location assigned to another location.

**Exception(s):**

2a. Database error.

2a1. System displays an error message

4a. User didn’t select any in DB “Location”.

4a1. System displays an error message and asks user to select records in DB “Location”.

5a. User didn’t select location to which selected location(s) should be assigned.

5a1. System displays an error message and asks user to select location to which selected location(s) should be assigned.

## 3.2.1.15. View/print plan of big location

**Actor(s)**: Users with appropriate permission.

**Precondition(s)**:

1. Authenticated session.
2. Necessity to view plan of big location.

**Trigger**: Clicking menu “Location ➔ View plan of big location”.

**Procedure:**

1. User on the page “Main” clicks on menu “Location ➔ View plan of big location” and system transfers in to page “Location ➔ View plan of big location”.
2. System displays list of locations for which plans are available.
3. User selects location which he/she wants to see the plan.

4. User clicks on the button “Show plan”.
5. System transfers user to page “Location ➔ View plan of big location➔ Perform” and shows plan of selected location.
6. User clicks on the picture of particular location.
   6.1. System shows to which Faculty, Department location belongs.
   6.2. System shows capacity of location (if applicable).
   6.3. System shows person/location assigned to this location (if applicable).
7. User chooses to zoom or/and print the plan.
8. User clicks on button “Close” in order to close plan.
9. System updates the action in Audit file.

**Post condition(s):**

1. Plan of location was viewed / printed.

**Exception(s):**

4a. User didn’t select location for which he/she wants to see the plan.

4a1. System displays an error message and asks person to select location.

## 3.2.1.16. Add new role (package of permissions)/permission

**Actor(s)**: Users with appropriate permission.

**Precondition(s)**:

1. Authenticated session.
2. Appropriate request is submitted and approved.
3. No such role/permission in the system.

**Trigger**: Clicking menu “Administration ➔ Add new role/permission”.

**Procedure:**

1. User on the page “Main” clicks on menu “Administration ➔ Add new role/permission” and system transfers to page “Administration ➔ Add new role/permission”.
2. User writes name of new role/permission.

3. User selects permissions (clicks on the checkboxes) for current role.
4. User terminates operation:
   4.1. User clicks on the button “Submit” to save changes.
      4.1.1. System automatically updates the changes in the audit file.
   4.2. User clicks on the button “Cancel” to abort changes.

**Post condition(s):**

1. New role/permission added to the system.

**Exception(s):**

4.1a. Compulsory field(s) is (are) empty.

4.1a1. System displays an error message and asks user to repeat again.

4.1b. User didn’t select any permission.

4.1b1. System displays an error message and asks user to select permissions.

4.1c. User wrote name of role which already exists.

4.1c1. System displays an error message and asks user to change name of role.

## 3.2.1.17. Edit role/permission

**Actor(s)**: Users with appropriate permission.

**Precondition(s)**:

1. Authenticated session.
2. Appropriate request is submitted and approved.
3. Necessity to change list of permissions for a particular requester.
4. Necessity to change name of permission.

**Trigger**: Clicking menu “Administration ➔ Edit role/permission”.

**Procedure:**

1. User on the page “Main” clicks on menu “Administration ➔ Edit role/permission” and system transfers to page “Administration ➔ Edit role/permission”.

2. System displays DB “Person” in a table. Displayed information depends on role, level, Faculty, and Department to which the user belongs.
3. User selects a requester clicking on appropriate checkbox.
4. User clicks button “Edit role/permission”.
5. Correction of role/permission.
   5.1. Correct role
      5.1.1. System displays a table, with a list of all available permissions to the right and a list of requester’s existing permissions to the left.
      5.1.2. User uses buttons➔, 🡐 in order to move permissions between two columns.
      5.1.3. User gives a due date for each permission added. If the user does not specify a date, the permission exists until the expiry of the account.
   5.2. Correct permission
      5.2.1. User correct name of permission.
6. User terminates operation:
   6.1. When user is satisfied with the list of permissions for requester, he/she clicks on button “Confirm”. The system updates the changes in the audit file.
   6.2. User aborts changes by clicking on button “Cancel”.

**Post condition(s):**

1. List of permissions for particular requester is changed. / Name of permission changed.

**Exception(s):**

2a. Database error.

2a1. System displays an error message

3a. User didn’t select record in DB “Person”.

3a1. System displays an error message and asks user to select record in DB “Person”.

3b. User selects more than one record in DB “Person”.

3b1. System displays an error message and asks user to select only one record in DB “Person”.

5.1a. List of requester’s permission is empty.

5.1a1. System displays an error message and asks user to select permissions for requester.

## 3.2.1.18. Assign role / permission to person(s) (bulk operation included)

**Actor(s)**: Users with appropriate permission.

**Precondition(s)**:

1. Authenticated session.
2. Appropriate request is submitted and approved.

**Trigger**: Clicking menu "Administration ➔ Assign role/permission to person(s)".

**Procedure:**

1. User on the page "Main" clicks on menu "Administration ➔ Assign role/permission to person(s)" and system transfers to page "Administration ➔ Assign role/permission to person(s)".
2. System displays DB "Person" in the table. Displayed information depends on role, level, Faculty, and Department to which the user belongs.
3. User uses scroll bar in order to see different records in the table.
4. User selects person(s) to which role/permission should be assigned by clicking on checkbox (es).
5. User selects role/ permission (from available list) to which person(s) should be assigned.
6. User terminates operation:

   6.1. User clicks on "Confirm" button to assign person(s) to role.

      6.1.1. System automatically updates the changes in the audit file.

   6.2. User clicks on "Cancel" button to abort operation.

**Post condition(s):**

1. Pole/permission assigned to person(s).

**Exception(s):**

2a. Database error.

2a1. System displays an error message.

6.1a. User didn't select any person in DB "Person".

6.1a1. System displays an error message and asks user to select person(s) in DB “Person”.

6.1b. User didn’t select role/permission to which selected person(s) should be assigned.

6.1b1. System displays an error message and asks user to select role/permission to which selected person(s) should be assigned.

### 3.2.1.19. Provide biometrical characteristic

**Actor(s)**: User with appropriate permissions.

**Precondition(s)**:

1. Authenticated session.
2. The user is a high privileged user.
3. The user does not have a biometric sample in the system.

**Trigger**: Clicking menu “Person ➔Provide biometrical characteristic”.

**Procedure:**

1. User on the page “Main” clicks on menu “Person ➔ Provide biometrical characteristic” and system transfers in to page “Person ➔Provide biometrical characteristic”.
2. System opens account of the user logged into system.
3. User on the tab “Authentication information” clicks on the button “Record voice”.
4. User provides sample of voice.
5. System records user’s voice into file, which will be used for additional personality check.
6. User clicks the button “Submit” to save sample.
7. System updates the action in Audit file.

**Post condition(s):**

1. High privileged user has biometrical characteristic (voice simple) to login into system.

**Exception(s):**

2a. Database error.

2a1. System displays an error message

3a. User didn’t click button “Record voice”.

3a1. System displays an error message and asks user to click button “Record voice”.

5a. Problems with saving file.

5a1. System displays an error message and asks user to record voice again.

6a. User didn’t click button “Submit”.

6a1. System displays an error message and asks user to click button “Submit”.

## 3.2.1.20. View “My profile”

**Actor(s)**: All users.

**Precondition(s)**:

1. Authenticated session.
2. Necessity to view “My profile”.

**Trigger**: Clicking menu “Reports➔“My profile”/”Asset➔“My profile”/”License➔“My profile”/ ”Location ➔“My profile”/ ”Administration ➔“My profile”.

**Procedure:**

1. User on the page “Main” clicks on menu “Reports➔“My profile”/”Asset➔“My profile”/”License➔“My profile”/ ”Location ➔“My profile”/ ”Administration ➔“My profile”.
2. User views list of assets/license assigned to him/her; list of assets/license borrowed by him/her; list of locations assigned to him/her; list of roles/permission, which he/she has.
3. User, if chooses, prints profile.

**Post condition(s):**

1. User viewed/printed own profile.

**Exception(s):**

1a. Database error.

1a1. System displays an error message

## 3.2.1.21. Add new request

**Actor(s)**: All users.

**Precondition(s)**:

1. Authenticated session.

**Trigger**: Clicking menu "Request➔Add new".

**Procedure:**

1. User on the page "Main" clicks on menu "Request➔Add new" and system transfers to page "Request➔Add new".
2. User selects kind of request: report a problem (reparation); report a bug (reparation); make a general request (acquisition, elimination); make a request to move item(s).
3. User chooses:
   - 3.1. User selects report a problem/ bug / general request.
     - 3.1.1. User types a request on the available textbox.
   - 3.2. User selects request move item(s).
     - 3.2.1. In the available field user types barcode(s) of item(s), which he/she wants to move.
     - 3.2.2. In appropriate list user selects the number of room, where he/she want to move selected item(s). Displayed list of locations depends on role, level, Faculty, and Department to which the user belongs.
     - 3.2.3. User types description of request in the textbox.
4. User clicks on "Submit" to submit request.
5. System verifies if asset(s)/license(s) belong to Faculty/Department to which the user belongs (for request move items).
6. System updates the request in Audit file.

**Post condition(s):**

1. System saves request in the DB and transfers for approval/rejection to person one level above.
2. If user with level3 requests, it is approved automatically.

**Exception(s):**

2a. User didn't select any kind of request.

2a1. System displays an error message and asks user to select one kind of request.

4a.User didn't write request in the textbox (for report the problem/ bug report/ general request).

4a1. System displays an error message and asks user to write request.

4b.User didn’t write barcode of item(s) (for request move item(s)).

4b1. System displays an error message and asks user to write barcode of item(s).

4c.User didn’t select location (for request move item(s)).

4c1. System displays an error message and asks user to select location.

4d. Item(s) doesn’t belong to the same Faculty/Department as user. (User doesn’t have right to move this asset).

4.1d1. System displays an appropriate message and tells that user doesn’t have right to move this asset.

### 3.2.1.22. Approve/Reject request

**Actor(s)**: Users with appropriate permission.

**Precondition(s)**:

1. Authenticated session.
2. There is/are request(s) in the system that needs to be approved/rejected.

**Trigger**: User clicks on a particular request, in the list of requests.

**Procedure:**

1. User on the page “Main” clicks on menu “Request➔Approve/Reject” and system transfers in to page “Request➔Approve/Reject”.
2. User views a list of requests that awaits a decision.
3. User selects a particular request from the list of requests.
4. The user views the requested text sent by the requesting user with the details of the user and user account.
5. User chooses a button according to the decision:

    4.1. Clicks on the “Approve” button that is found below the request text, if the user approves the request.

    4.1.1. System automatically updates the changes in the audit file.

    4.2. Clicks on the “Reject” button that is found below the request text, if the user rejects the request. The user types in a reason for rejection.

6. The system saves the decision and updates the audit file.
7. The system maps the request to the next level, if the request is approved.
8. The system sends an email to the requested user, if the request is rejected

**Post condition(s):**

1. The request has been approved or rejected.

**Exception(s):**

5.2a. User didn't write reason for rejection request.

5.1a1. System displays an error message and asks user to write reason for rejection request.

## 3.2.1.23. View list of all requests in the system

**Actor(s)**: Users with appropriate permission.

**Precondition(s)**:

1. Authenticated session.

**Trigger**: Clicking menu "Request➔View list of all requests".

**Procedure:**

1. User on the page "Main" clicks on menu "Request➔View list of all requests" and system transfers in to page "Request➔View list of all requests".
2. The system shows a list of all requests on the screen.
3. The user clicks on a request of choice.

**Post condition(s):**

1. Request(s) was(were) viewed by user.

**Exception(s):**

No exceptions.

## 3.2.1.24. Add new Faculty/Department

**Actor(s)**: Users with appropriate permission.

**Precondition(s)**:

1. Authenticated session.

2. Appropriate request is submitted and approved.
3. No such Faculty/Department in the DB.

**Trigger**: Clicking menu “Faculty/Department ➔ Add new”.

**Procedure:**

1. User on the page “Main” clicks on menu “Faculty/Department ➔ Add new” and system transfers to page “Faculty/Department ➔ Add new”.
2. User provides the name, type, and building of Faculty/Department.
3. System automatically generates the Faculty/Department created date.
4. User terminates operation:

   4.1. User clicks on the “Submit” button to save changes.

      4.1.1. System automatically updates the changes in the audit file.

   4.2. User clicks on the “Cancel” button to abort changes.

**Post condition(s):**

1. New Faculty/Department has been added to the system.

**Exception(s):**

4.1a Compulsory field(s) is (are) empty.

4.1.a1 System displays an error message and asks user to repeat again.

## 3.2.1.25. Basic search

**Actor(s)**: All users.

**Precondition(s)**:

1. Authenticated session.

**Trigger**: Clicking on “Search” in the menu.

**Procedure:**

1. On the “Main” page, user clicks on “Search” and the system transfers to “Search” page.
2. The page displays a textbox, a “Search” button, and an “Advanced Search” link.

   2.1. Basic search.

2.1.1. The user types his/her query in the textbox.

2.1.2. The user clicks on the “Search” button or hits the enter key.

2.1.3. The system reads the query as a case insensitive substring.

2.1.4. The system matches the substring with the records in the database.

2.1.5. The system displays the corresponding matches in the form of a table.

2.2. Advanced search.

2.2.1. The system redirects to the “Advanced Search” page (see next Use Case).

**Post condition(s):**

1. System displays search result.

**Exception(s):**

2.1.1a. User didn’t type anything in the textbox for search.

2.1.1a1. System displays an error message and asks user to type a query for search.

2.1.4a. Database error.

2.1.4a1. System displays an error message.

2.1.4b. Item not found.

2.1.4b1. System displays “Search item not found” message and prompts the user to change the query for search.

### 3.2.1.26. Advanced search

**Actor(s)**: All users.

**Precondition(s)**:

1. Authenticated session.

**Trigger**: Clicking on the link “Advanced search” in the “Search” page.

**Procedure:**

**1.** On the “Main” page, user clicks on “Search” and the system transfers to “Search” page.

2. The user then clicks on the link “Advanced search” and the system transfers to “Advanced search” page.

3. The advanced search page contains a textbox for the user to type the query, a search button, a link “Basic search” that transfers to basic search page, and a table with four columns. The columns show the categories in the database, named as Persons, Locations, Assets, and Licenses, with a box and a list for each category.

4. Select region of search.

   4.1. The user restricts the search fields by selecting one or more categories on clicking the appropriate box (es), which the system marks with a checkmark.

   4.2 User can further restrict the search by selecting one or more fields from the lists under the required category, by clicking on the name of the fields.

5. Type query for search.

   5.1. User types a simple query (one string without combinations) for search.

   5.2. User types a complex string (two or more strings with AND/OR/NOT or their combinations) for search.

6. The user clicks on the “Search” button or hits the enter key.

7. The system reads the string or strings as a case insensitive substring(s).

8. The system matches the string(s) with the records in the database.

9. The system displays the corresponding matches in the form of a table.

**Post condition(s):**

1. System displays search result.

**Exception(s):**

8a. User didn’t type anything in the textbox for search.

8a1. System displays an error message and asks user to type a query for search.

8b. User typed an incorrect string in the textbox for search (for example, string can’t start with AND, OR, etc.).

8b1. System displays an error message and asks the user to retype string for search.

9a. Database error.

9a1. System displays an error message.

9b. Item not found.

9b1. System displays “Search item not found” message and prompts the user to change the query for search.

### 3.2.1.27. Login

**Actor(s)**: All users.

**Precondition(s)**:

1. Unauthenticated session.

**Trigger**: Clicking on Login textbox in the home page.

**Procedure:**

1**.** The user clicks on the Login textbox in the home page.

2. The user types in the username (eight characters beginning with a letter) and password and clicks on “Submit” or just hits the enter button.

3. The system verifies the username and password.

3.1. The system opens the authenticated session, if user information was true.

3.2. The system displays the login page with “Login Failure” information, if user information is false.

4. If the user has more than one role, the system gives the user’s roles in a list.

5. The user picks a role for the session.

6. If the user is a high-privileged user, the system prompts user to record his/her voice.

7. System updates the information in Audit file.

**Post condition(s):**

1. The user is logged in to the system (Authenticated session).

**Exception(s):**

3a. Database error.

3a1. System displays an error message.

3b. Invalid Username/Password.

3b1. System displays error message and prompts user to try again.

6. System displays error message, if the user's voice is not recognized, and prompts user to retry.

### 3.2.1.28. Logout

**Actor(s)**: All users.

**Precondition(s)**:

1. Authenticated session.

**Trigger**: The user clicks on "Logout" in any page (or the session expires automatically if unused for 30 minutes).

**Procedure:**

1. The user clicks on the "Logout" button located on any page in the authenticated session.

2. The system terminates the session.

3. The system transfers to home page and displays "Logged out successfully".

4. System updates the information in Audit file.

**Post condition(s):**

1. Unauthenticated session.

**Exception(s):**

1. System terminates the session if unused for 30 minutes.

### 3.2.1.29. Create/Print reports

**Actor(s)**: Users with appropriate permission.

**Precondition(s)**:

1. Authenticated session.

**Trigger**: Clicking on menu "Report" in the main page.

**Procedure:**

1. User on the page "Main" clicks on menu "Reports" and system transfers to page "Reports".
2. User selects type of location from available list: research lab, teaching lab, office.

3. User selects kind of report for chosen type of location: compare capacity with number of chairs; compare capacity with number of tables; compare capacity with number of PC (for teaching lab and offices only); compare capacity with number of students (for research lab only).
4. User clicks button “Create report”.
5. Report will be displayed in a table format.
6. Report contains columns: location number, capacity of location, number of chairs/tables/PC/students assigned to location, and difference between two last columns.
7. User, if prefers, sorts data by number of location and prints report.

**Post condition(s):**

1. System displays report result.
2. User viewed/printed report results.

**Exception(s):**

4a. User didn’t select type of location.

4a1. System displays an error message and asks user to select type of location.

4b. User didn’t select kind of report.

4b1. System displays an error message and asks user to select kind of report.

### 3.2.1.30. Auditing

**Actor(s)**: Users with ‘audit’ permission.

**Precondition(s)**:

1. Authenticated session.

**Trigger**: Clicking on menu “Auditing” in the main page.

**Procedure:**

1. User on the page “Main” clicks on menu “Auditing” and system transfers in to another “Login page for “Auditing”.
2. After login procedure, if the user is authenticated, the system transfers to “Auditing” page.
3. User selects a specific period of time for which he/she wants to see records of all actions done in the system.

4. Users makes more choices:
    4.1. User does not specify additional parameters.
    4.2. User specifies additional parameters.
        4.2.1. User clicks on “Select person(s)” button.
            a. In DB “Person” user selects person(s) clicking on appropriate checkbox (es).
            b. User clicks on “Continue” button.
        4.2.2. User clicks on “Select item(s)” button.
            a. In DB “Asset”/”License” user selects person(s) clicking on appropriate checkbox (es).
            b. User clicks “Continue” button.
5. User clicks on “Create report” button.
6. Report is displayed in a table format.
        6.1.1. System creates one big report of all actions taken place in the system, if the user does not specify constraints.
        6.1.2. System short lists the report if the user provides constraints.
7. User, if prefers, prints the report.

**Post condition(s):**

1. System displays report result.
2. User viewed/printed report results.

**Exception(s):**

6. a. Database error.

6. a1. System displays an error message.

## 3.2.1.31. Help

**Actor(s)**: All users.

**Precondition(s)**:

1. User is in any of the UUIS web page.

**Trigger:** User clicks on the link ‘Help’ in any of the UUIS web page.

**Procedure:**

1. User clicks on one of the ‘Help’ link available in all pages of UUIS.
2. A new window opens.
3. The window displays all the functions that could be performed on the page where the help was clicked.
4. The user clicks on the function that he/she needs help with.
5. The system displays the details of the chosen function and guides the user to operate the function.
6. The user clicks on the “Close” button of the help window to come back to the original window the user was working on.

**Post condition(s):**

The user is back on the page where he/she clicked on ‘Help’ link.

**Exception(s):**

No exceptions.

### 3.2.2. Use case diagrams

The following are use case diagrams, each diagram with few use cases combined together:

#### 3.2.2.1. Login, Provide biometrical characteristic, Select language, Help, & Logout

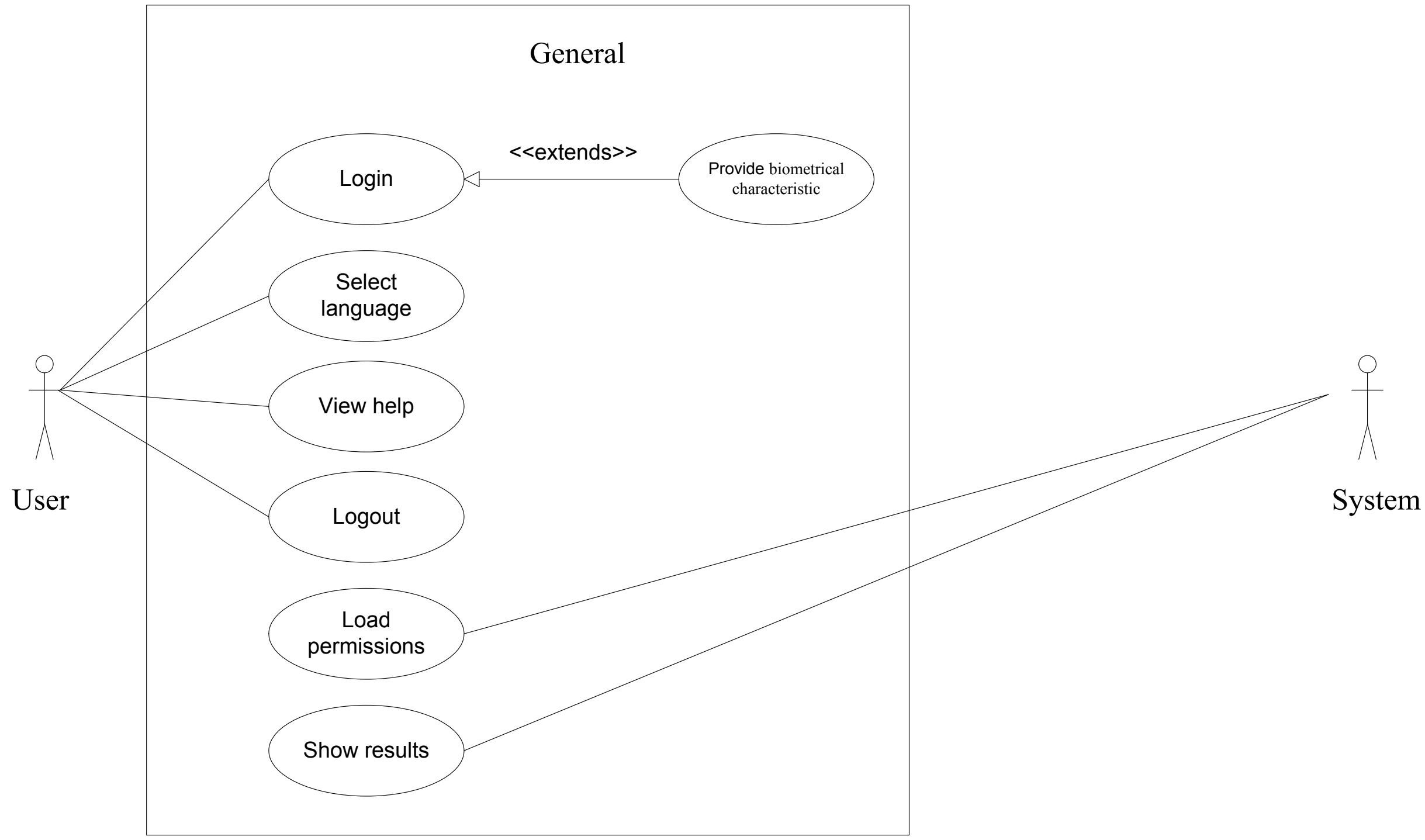

**Figure 1.** Use case diagram for Login, Provide biometrical characteristic, Select language, Help, & Logout

### 3.2.2.2. Create a group of assets / locations, Create new type of asset / license / location / person, & Create new subgroup of assets

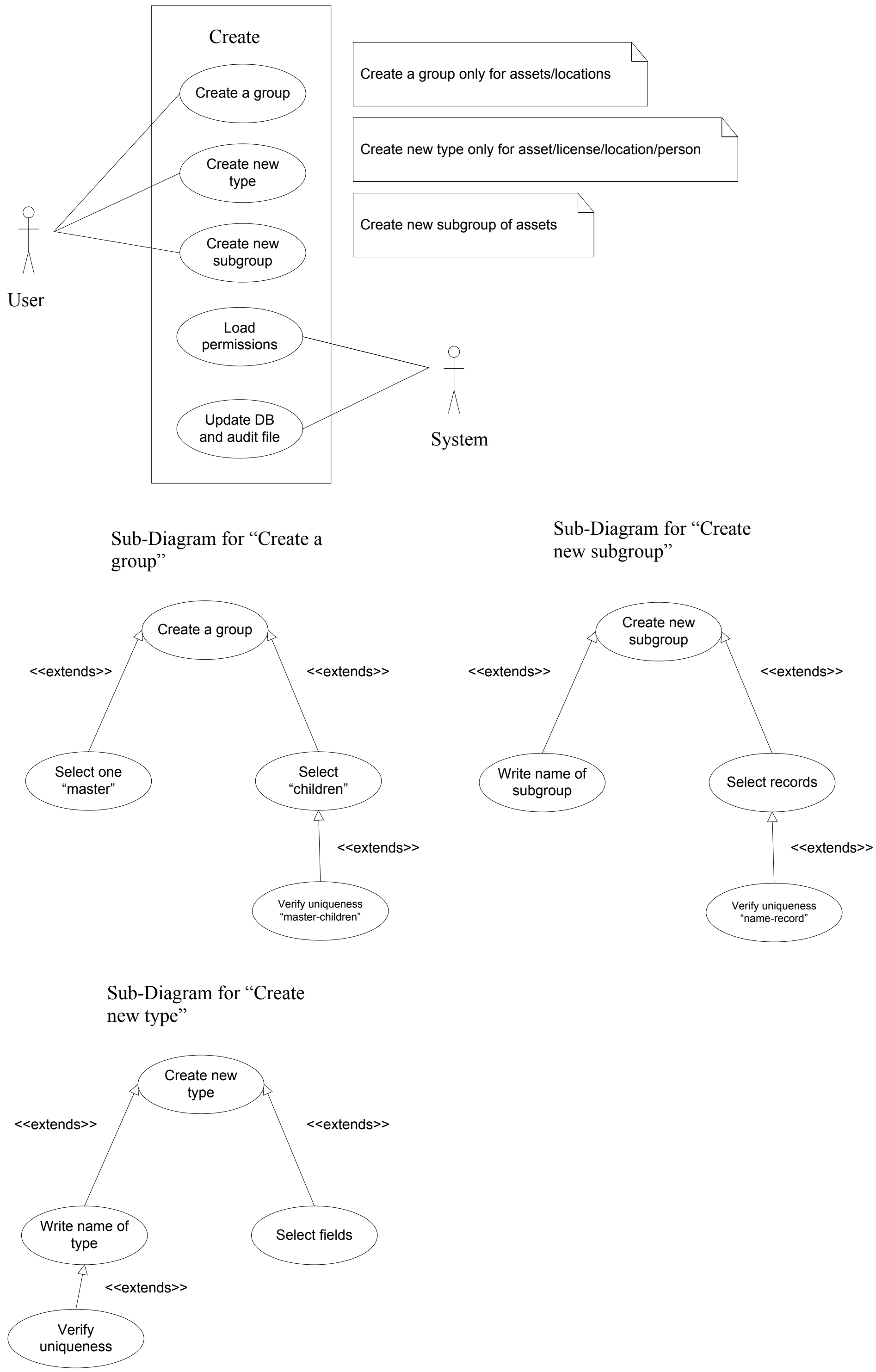

**Figure 2.** Create a group of assets / locations, new type of asset / license / location / person, & new subgroup of assets

### 3.2.2.3. Add new asset / license / location / faculty / department / role

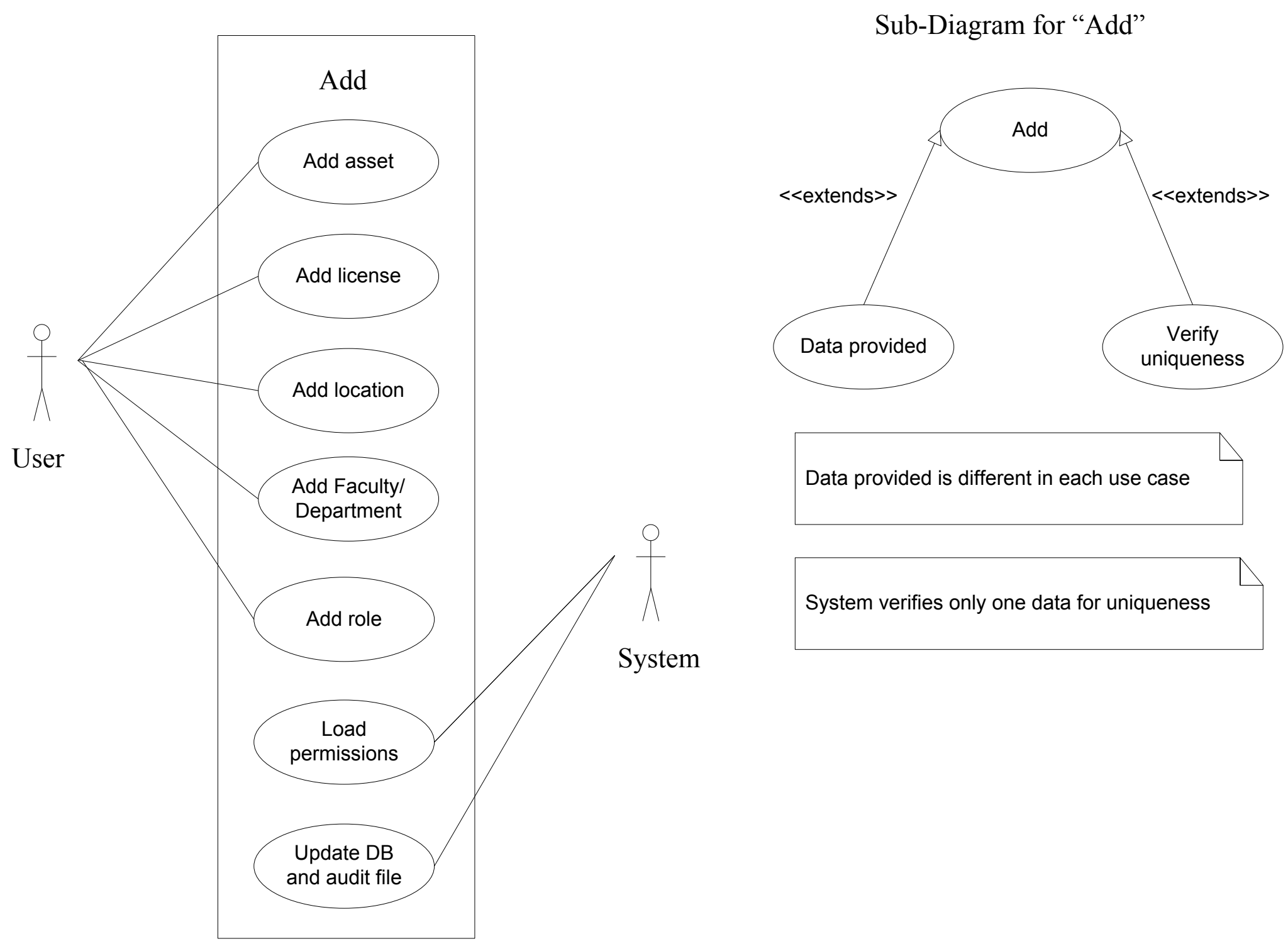

**Figure 3**. Add new asset / license / location / faculty / department / role

### 3.2.2.4. View DB “Asset” / DB “License” / DB “Location” / DB “Person” / Faculty / Department & View plan of big location

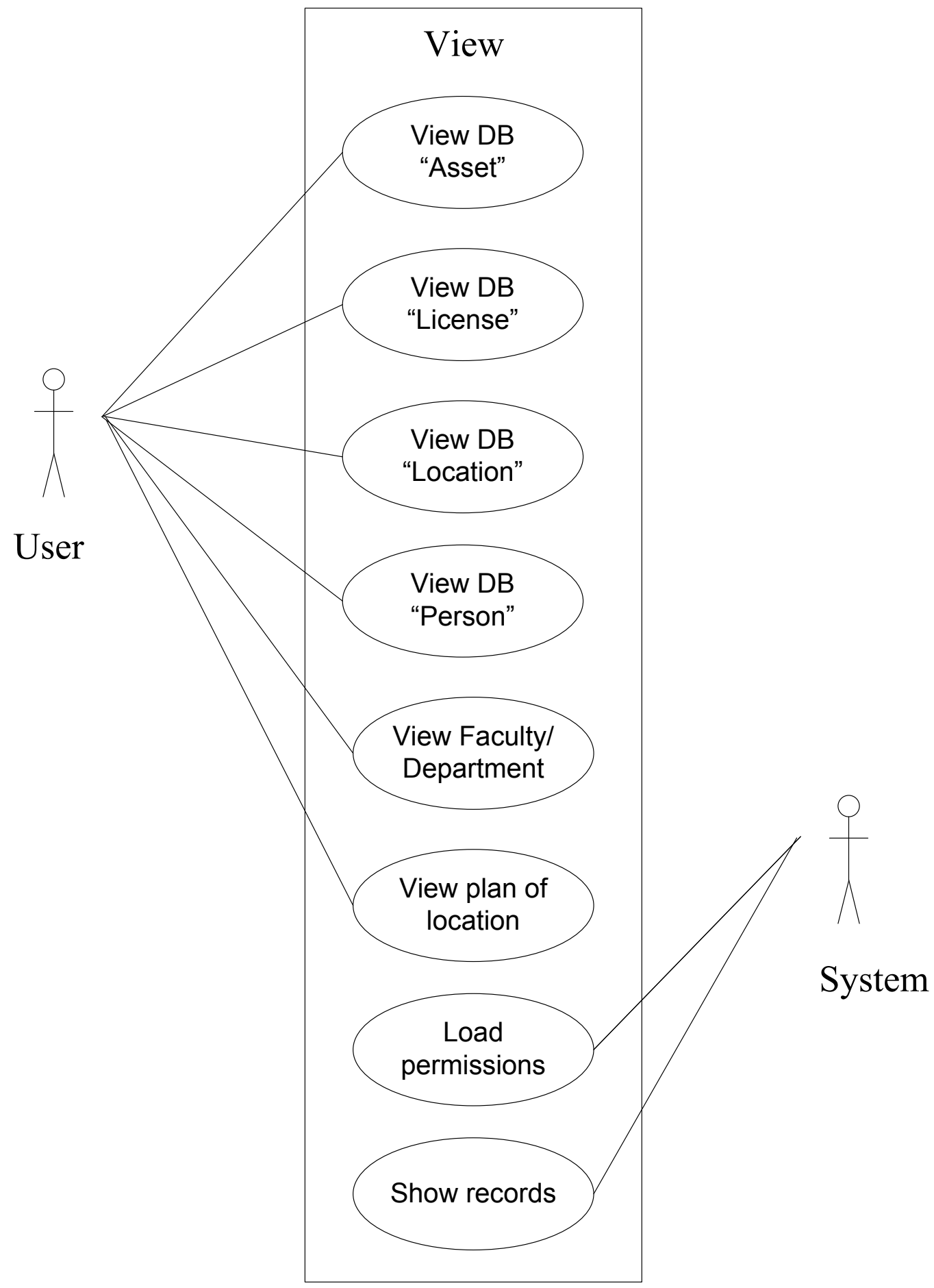

**Figure 4**. View DB “Asset” / DB “License” / DB “Location” / DB “Person” / Faculty / Department & View plan of big location

#### 3.2.2.5. Edit record(s) in DB “Asset” / DB “License” / DB “Location” / DB “Person” / Faculty / Department & Edit role / permission(s)

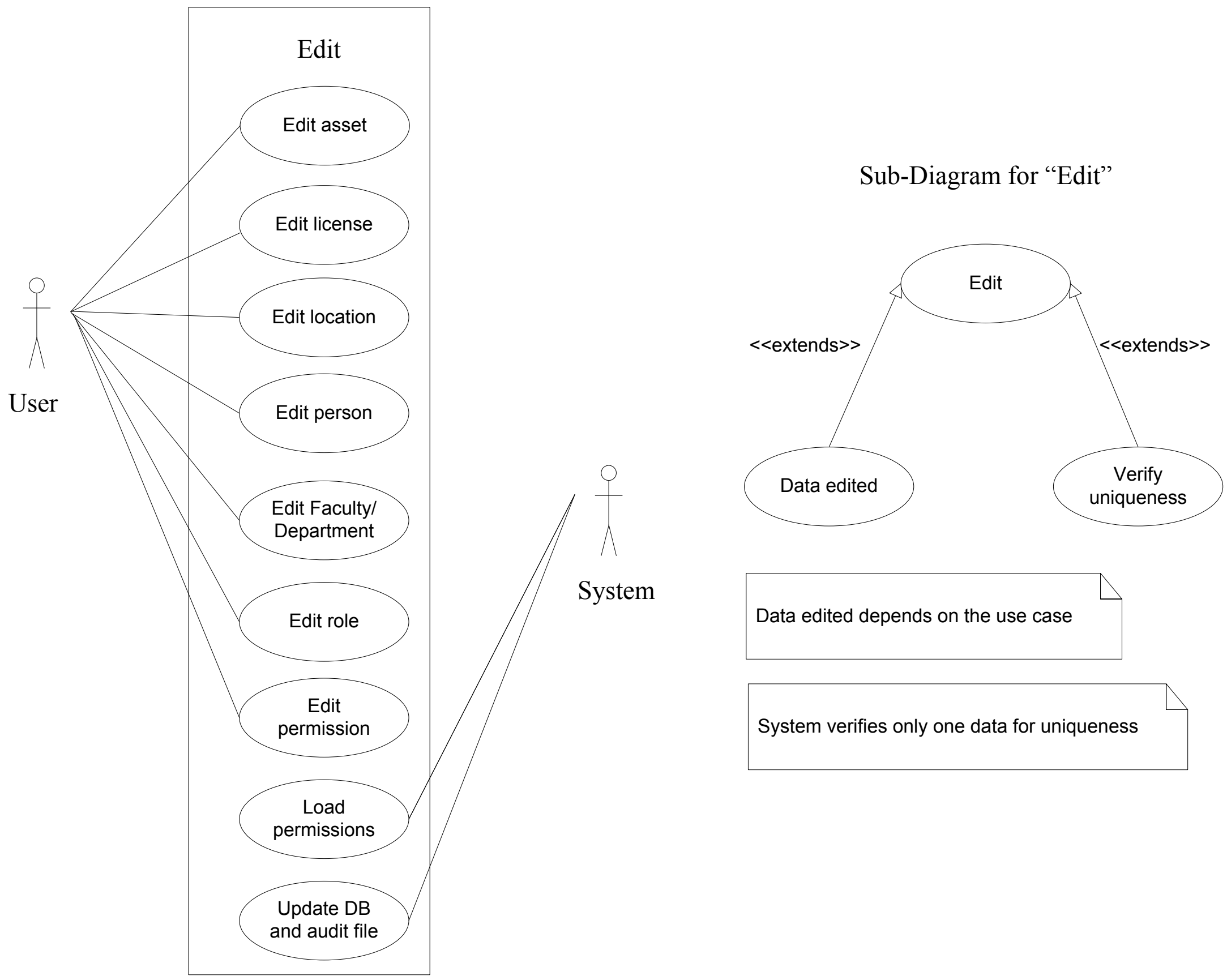

**Figure 5.** Edit record(s) in DB “Asset” / DB “License” / DB “Location” / DB “Person” / Faculty / Department & Edit role / permission(s)

### 3.2.2.6. Delete record(s) in DB “Asset” / DB “License” / DB “Location” / DB “Person”

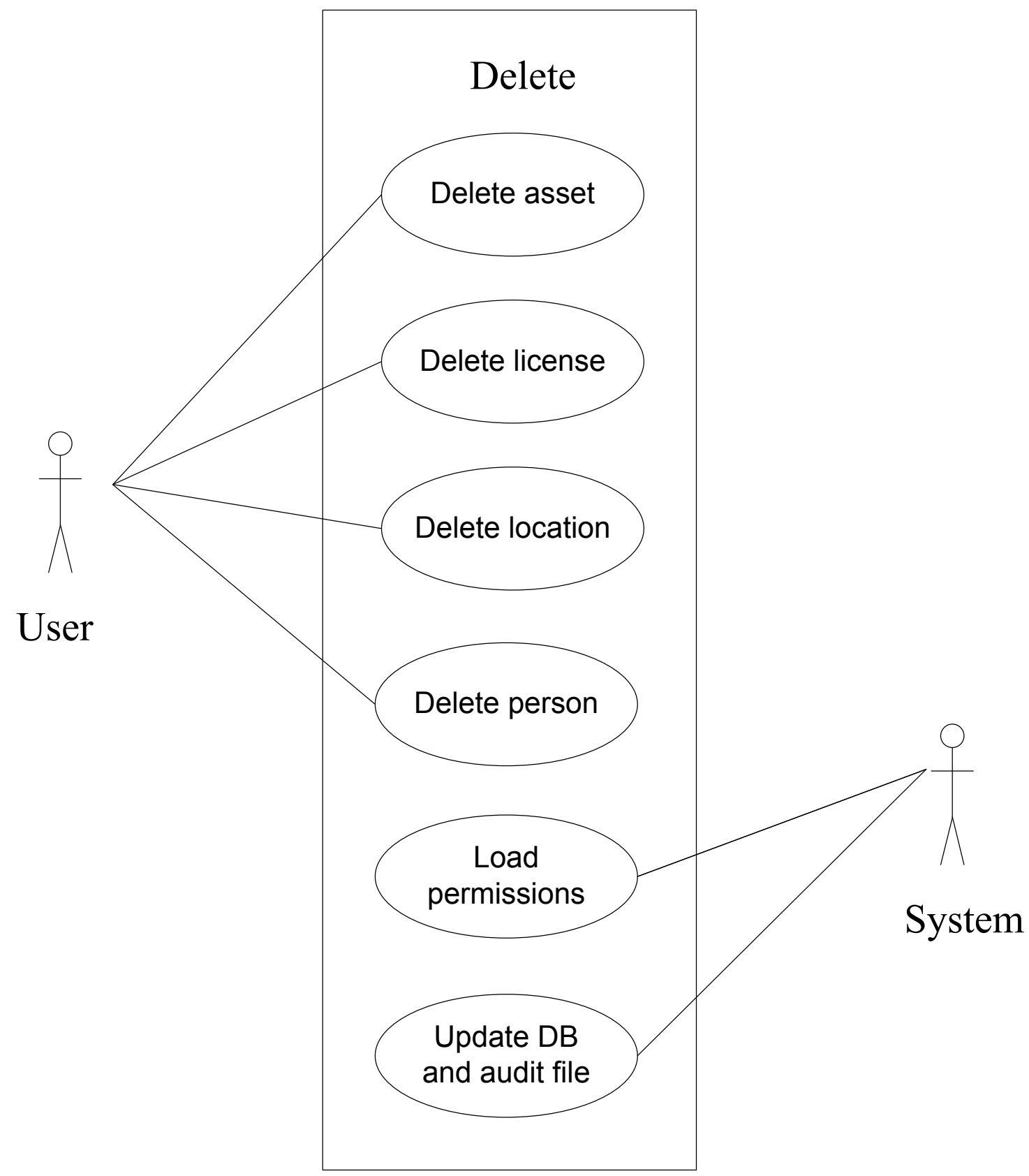

**Figure 6.** Delete record(s) in DB “Asset” / DB “License” / DB “Location” / DB “Person”

### 3.2.2.7. Import of assets / licenses / locations / persons (accounts)

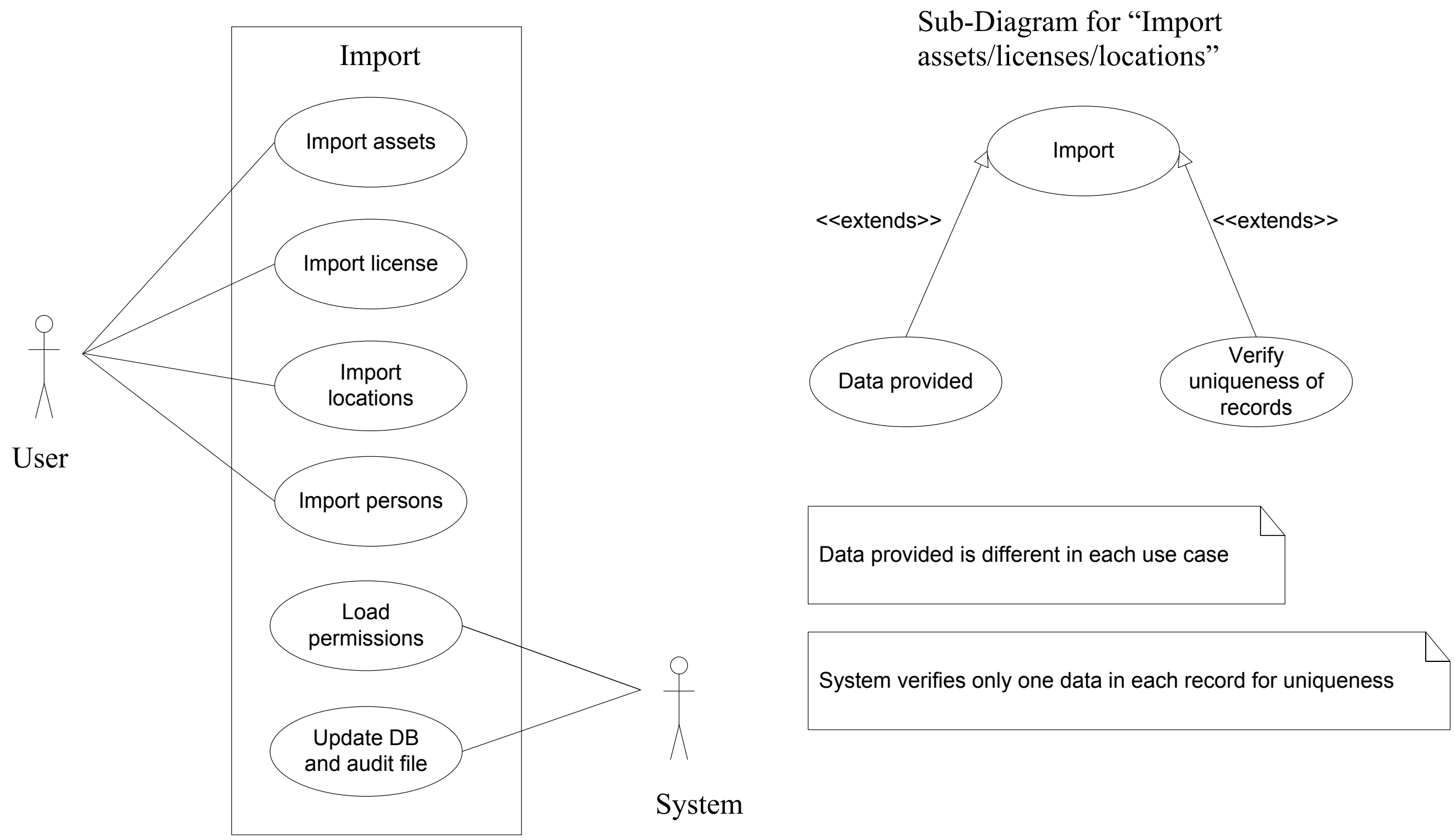

**Figure 7**. Import of assets / licenses / locations / persons (accounts)

### 3.2.2.8. Borrow asset(s) / license(s)

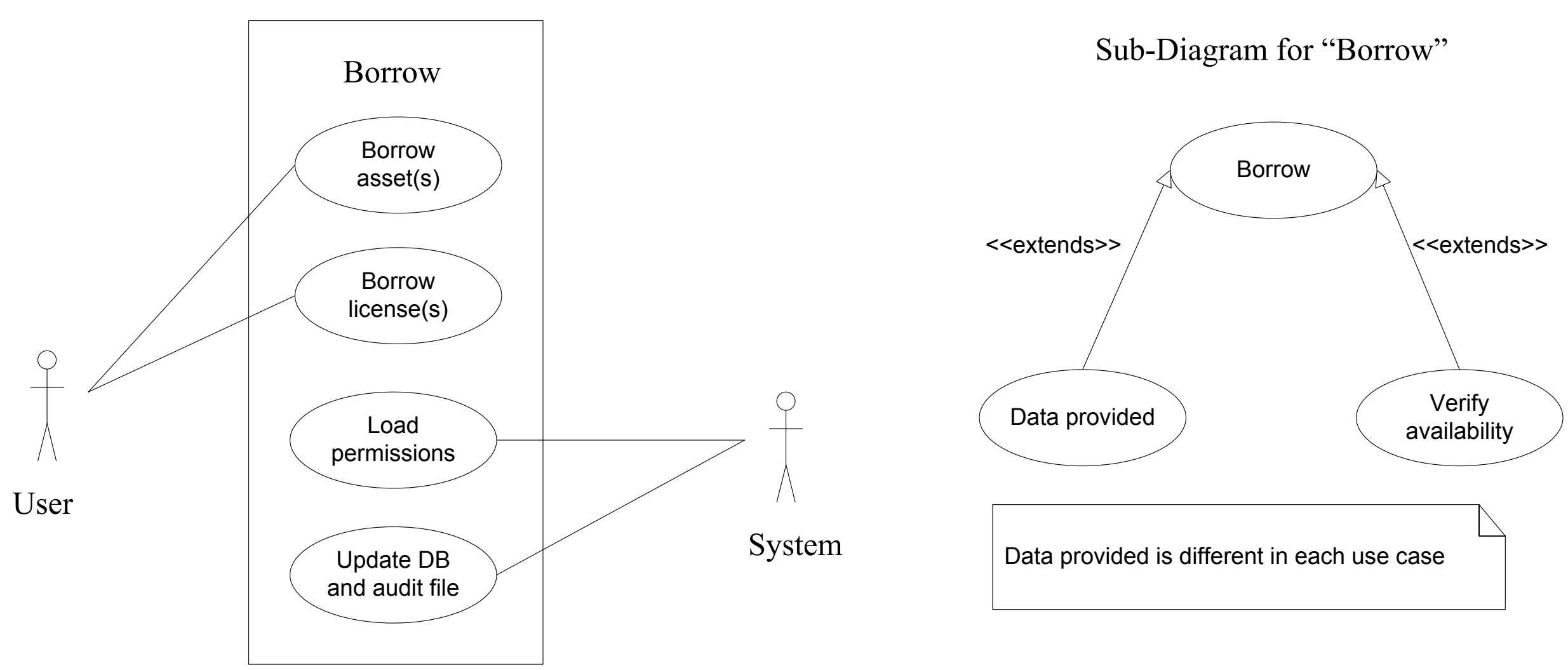

**Figure 8.** Borrow asset(s) / license(s)

## 3.2.2.9. Add new request, Approve / Reject request, & View list of all requests in the system

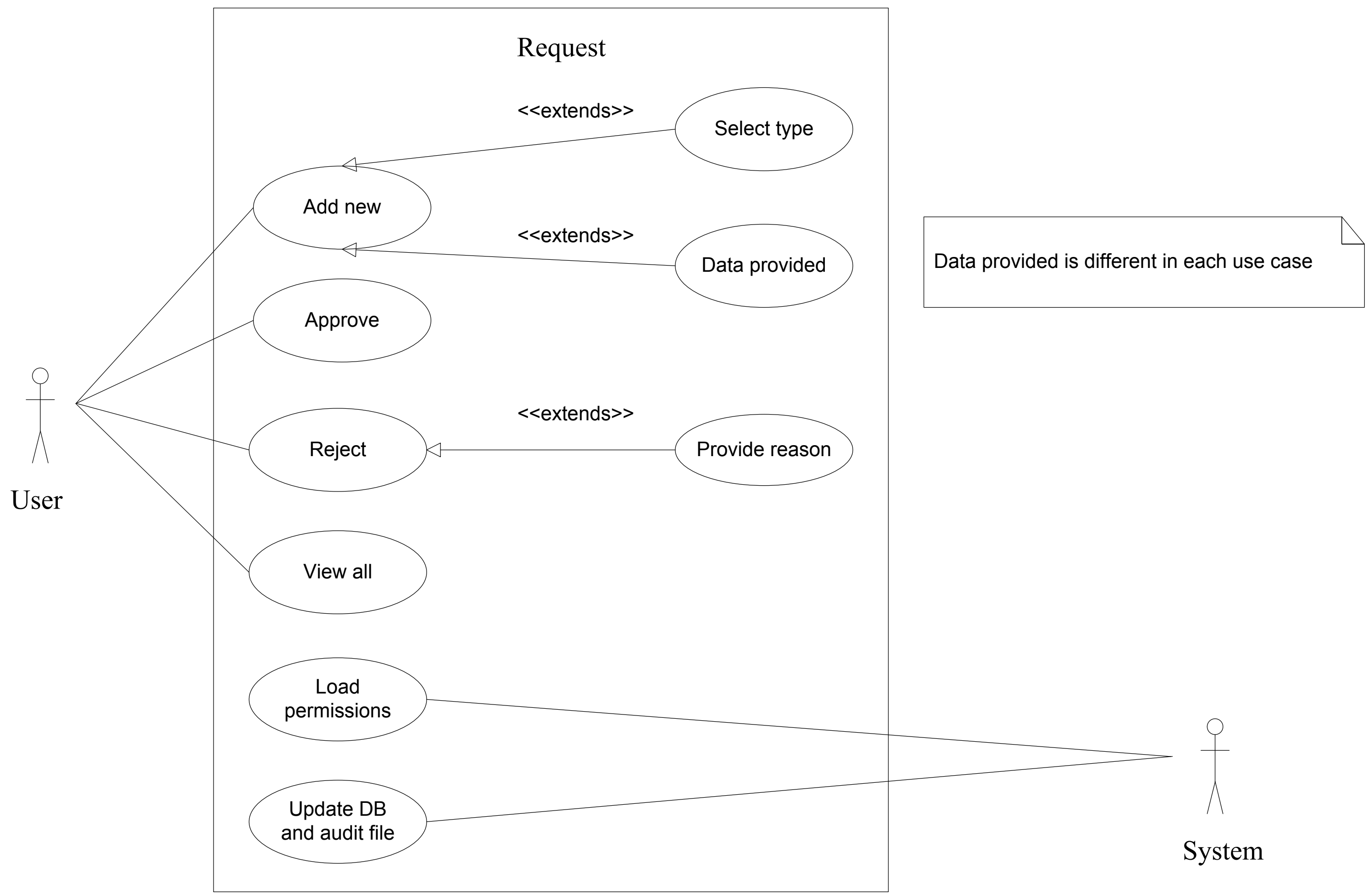

**Figure 9**. Add new request, Approve / Reject request, & View list of all requests in the system

### 3.2.2.10. Assign asset(s) to person / location, Assign license to asset, Assign location(s) to another location / Department, & Assign role / permission to person(s)

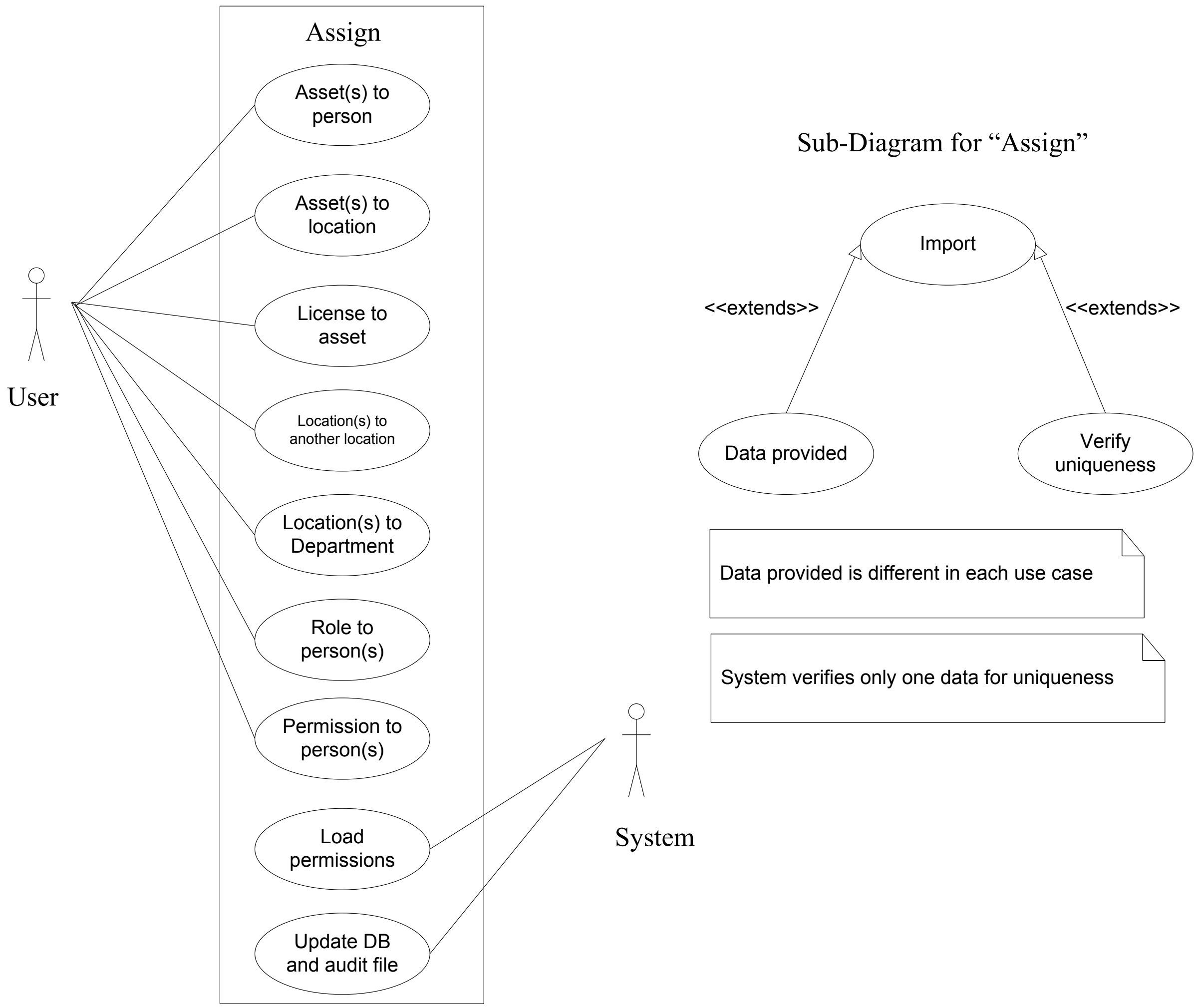

**Figure 10**. Assign asset(s) to person / location, Assign license to asset, Assign location(s) to another location / Department, & Assign role / permission to person(s)

### 3.2.2.11. Basic search & Advanced search

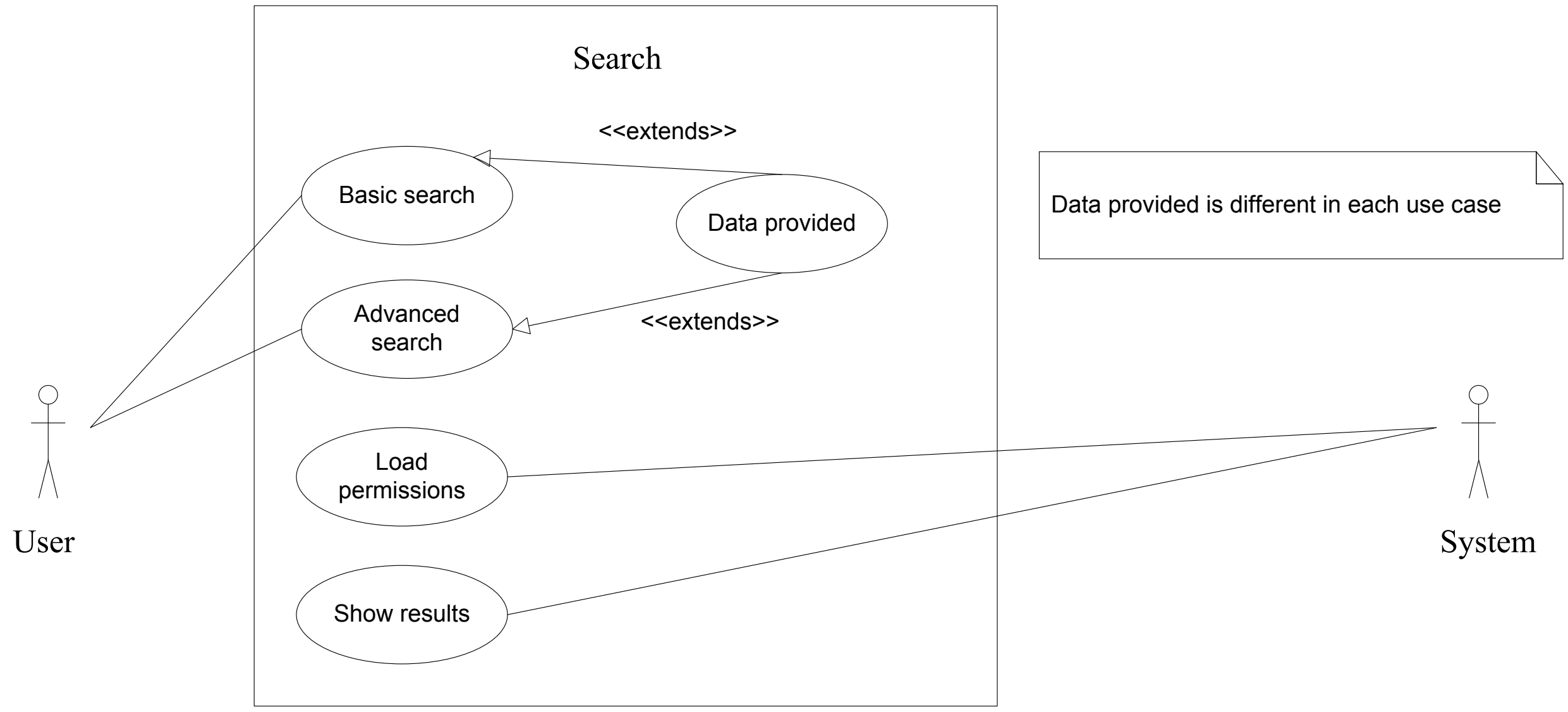

**Figure 11**. Basic search & Advanced search

## 3.2.2.12. View “My profile”, Create / print reports, & Auditing

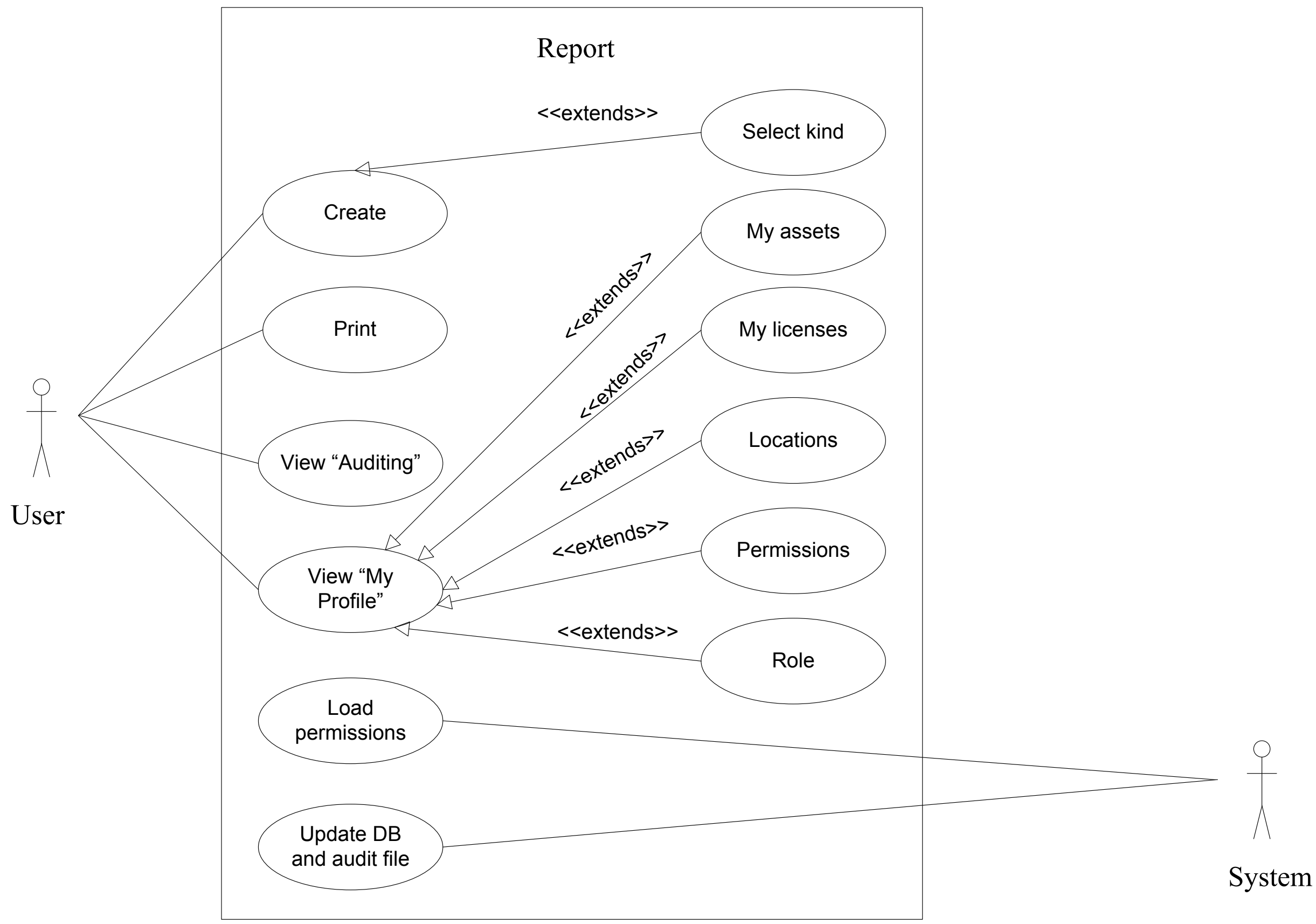

**Figure 12.** View “My profile”, Create / print reports, & Auditing

## 3.2.3. Data flow diagram

Data flow diagram for the project:

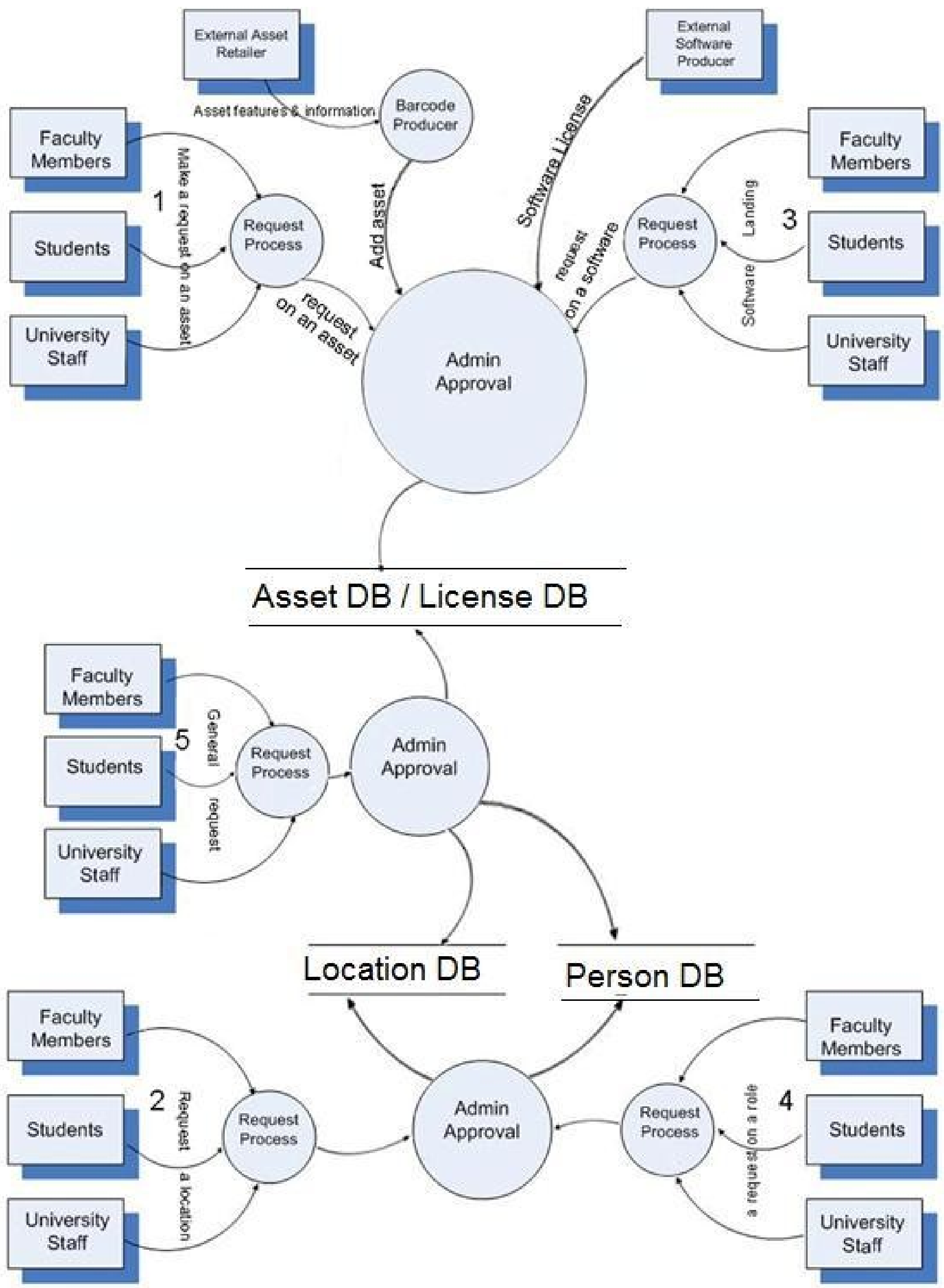

**1, 2, 3, 4, 5 – Please check under the figure for explanations**

**Figure 13.** Data Flow diagram

On Figure 13, the numbers 1, 2, 3, 4, 5 mean the following:

1 – insertAsset, seeAssets, editAsset, deleteAssets, borrowAssets, addGroupAsset, addTypeAsset, addSubgroupAsset, importAsset, assignAssetsToPerson, assignAssetsToLocation, seeMyAssets

2 – insertLocation, seeLocations, editLocation, deleteLocations, addGroupLocation, addTypeLocation, see_printFloorPlan, importLocation, assignLocationToPerson, assignLocationToLocation, assignLocationToDepartment, seeMyLocations

3 – insertLicense, seeLicenses, editLisense, deleteLicenses, borrowLicenses, addTypeLicence, importLicense, assignLicenceToAsset, seeMyLicenses

4 - seePersons, editPerson, deletePersons, addBiometric, importPerson

5 – addRole, editRole, addPermission, editPermission, assignPermissionToPersons, assignRoleToPersons, seeMyRole, seeMyPermissions, insertFacDep, seeFacDep, editFacDep, createAcquisitionRequest, createReparationRequest, createEliminationRequest, createMoveRequest, aprove_rejectRequest, seeRequestsAll, basicSearch, advancedSearch, create_printReport, seeAudit, seeMyProfile, selectLanguage, login_logout.

## 3.3. Performance requirements

The system is required to support multiple terminals simultaneously. The system should handle reasonable number of users without break or inconsistency.

## 3.4. Database requirements

ER diagram:

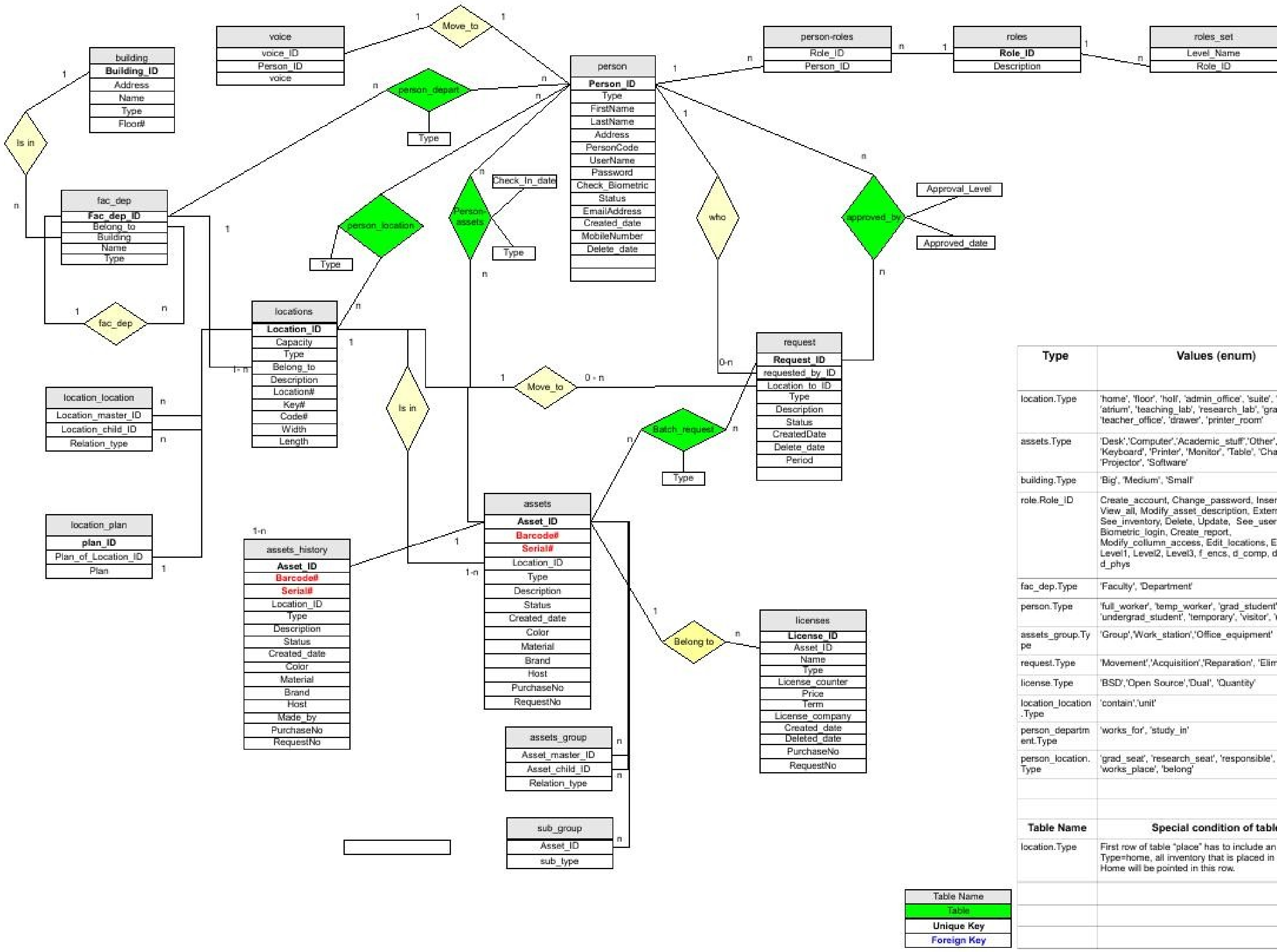

| Type | Values (enum) |
|---|---|
| location.Type | 'home', 'floor', 'holl', 'admin_office', 'suite', 'cubicle', 'atrium', 'teaching_lab', 'research_lab', 'grad_seat', 'teacher_office', 'drawer', 'printer_room' |
| assets.Type | 'Desk','Computer','Academic_stuff','Other', 'Mouse', 'Keyboard', 'Printer', 'Monitor', 'Table', 'Chair', 'Projector', 'Software' |
| building.Type | 'Big', 'Medium', 'Small' |
| role.Role_ID | Create_account, Change_password, Insert_into_DB, View_all, Modify_asset_description, External_acces, See_inventory, Delete, Update, See_users, Biometric_login, Create_report, Modify_collumn_access, Edit_locations, Edit_persons, Level1, Level2, Level3, f_encs, d_comp, d_math, d_phys |
| fac_dep.Type | 'Faculty', 'Department' |
| person.Type | 'full_worker', 'temp_worker', 'grad_student', 'undergrad_student', 'temporary', 'visitor', 'undefined' |
| assets_group.Type | 'Group','Work_station','Office_equipment' |
| request.Type | 'Movement','Acquisition','Reparation', 'Elimination' |
| license.Type | 'BSD','Open Source','Dual', 'Quantity' |
| location_location.Type | 'contain','unit' |
| person_department.Type | 'works_for', 'study_in' |
| person_location.Type | 'grad_seat', 'research_seat', 'responsible', 'works_place', 'belong' |
| | |
| | |

| Table Name | Special condition of table |
|---|---|
| location.Type | First row of table "place" has to include an row with Type=home, all inventory that is placed in People's Home will be pointed in this row. |
| | |
| | |
| | |

**Figure 14**. ER diagram

## 3.5. Quality requirements

### 3.5.1. Reliability

The system should work reliably, with automatic backup and recovery features. In case of unexpected termination of a session, the unsaved data should be recovered without loss and displayed to the respective users for saving into the system or continuing with the work. At any time, audit file and all db and mailing information are required to be updated in the backup.

### 3.5.2. Availability

The entire system should be available round the year, except for a periodic maintenance. The maintenance period should be pre scheduled and short. The users should be reminded of the unavailability period, well in advance.

### 3.5.3. Security

The system, at any time, should be accessed only by the authenticated users. Network communications should use cryptographic protocols such as SSL. Automated responses should be restricted using CAPTCHA. The system is required to end the session automatically, when an open session is not used for a specific period of time.

### 3.5.4. Maintainability

The document should be easy for the users who execute the system day to day, for the developers who wish to edit or develop further, and for the personnel who is in charge of the maintenance.

### 3.5.5. Portability

The system should support new versions of the related browsers. The administrative and server technologies should be standard and supported by most platforms.

### 3.5.6. Usability

The GUI should be easy to learn and use by users of any technical background. A built-in help feature should be available in all pages, to guide the users with the available functions on that page. An easy to understand documentation should be provided with the system. System should support several languages.

# 4. Appendix

## A. Cost estimation

## COCOMO model – Organic software mode

The model estimates cost using one of three different development modes: organic, semidetached, and embedded. Here is a summary of how Boehm describes the modes:

> "Organic
>
> In the organic mode, relatively small software teams develop software in a highly familiar, in-house environment. Most people connected with the project have extensive experience in working with related systems within the organization, and have a thorough understanding of how the system under development will contribute to the organizations objectives. Very few organic-mode projects have developed products with more than 50 thousand delivered source instructions (KDSI)".

**Table 4.** Cost estimation with COCOMO model, Organic software mode

| Ratings | | | | | | |
|---|---|---|---|---|---|---|
| | **Very Low** | **Low** | **Nominal** | **High** | **Very High** | **Extra High** |
| **Product attributes** | | | | | | |
| Required software reliability | 0.75 | 0.88 | 1 | **1.15** | 1.4 | |
| Size of application database | | **0.94** | 1 | 1.08 | 1.16 | |
| Complexity of the product | 0.7 | 0.85 | **1** | 1.15 | 1.3 | 1.65 |
| **Hardware attributes** | | | | | | |
| Run-time performance constraints | | | **1** | 1.11 | 1.3 | 1.66 |
| Memory constraints | | | **1** | 1.06 | 1.21 | 1.56 |
| Volatility of the virtual machine environment | | **0.87** | 1 | 1.15 | 1.3 | |
| Required turnabout time | | **0.87** | 1 | 1.07 | 1.15 | |
| **Personnel attributes** | | | | | | |
| Analyst capability | 1.46 | 1.19 | **1** | 0.86 | 0.71 | |
| Applications experience | 1.29 | **1.13** | 1 | 0.91 | 0.82 | |
| Software engineer capability | 1.42 | **1.17** | 1 | 0.86 | 0.7 | |
| Virtual machine experience | 1.21 | **1.1** | 1 | 0.9 | | |
| Programming language experience | 1.14 | **1.07** | 1 | 0.95 | | |
| **Project attributes** | | | | | | |
| Application of software engineering methods | 1.24 | 1.1 | **1** | 0.91 | 0.82 | |
| Use of software tools | 1.24 | 1.1 | **1** | 0.91 | 0.83 | |
| Required development schedule | 1.23 | 1.08 | 1 | **1.04** | 1.1 | |
| | | | | | | |
| **Effort applied, per/month** | | | | | | |
| **E = a * $(KLOC)^b$ *EAF** | | | | | | |
| EAF (effort adjustment factor) = | 1.32 | | | | | |
| KLOC (number of lines, thousands)= | 3.5 | | | | | |
| a (coefficient for Organic software model) = | 3.2 | | | | | |
| b (coefficient for Organic software model) = | 1.05 | | | | | |
| **E (efford applied, per/month)=** | **15.79** | | | | | |
| | | | | | | |
| **Development time, month** | | | | | | |
| **$D = c * (E)^d$** | | | | | | |
| c (coefficient for Organic software model) = | 2.5 | | | | | |
| d (coefficient for Organic software model) = | 0.38 | | | | | |
| **D (development time, month)=** | **7.13** | | | | | |
| | | | | | | |
| **People required, per** | | | | | | |
| **P = E / D** | | | | | | |
| **P (people required, per) =** | **2.21** | | | | | |
| | | | | | | |
| **Cost of project, dollars** | | | | | | |
| **C = P * D * CP** | | | | | | |
| CP (cost per person-month, dollars)= | 4800 | | | | | |
| **C (cost of project, dollars)** | **75788.0912** | | | | | |
| | | | | | | |

## COCOMO II model

This estimation was done, by using online service COCOMO II - Constructive Cost Model.

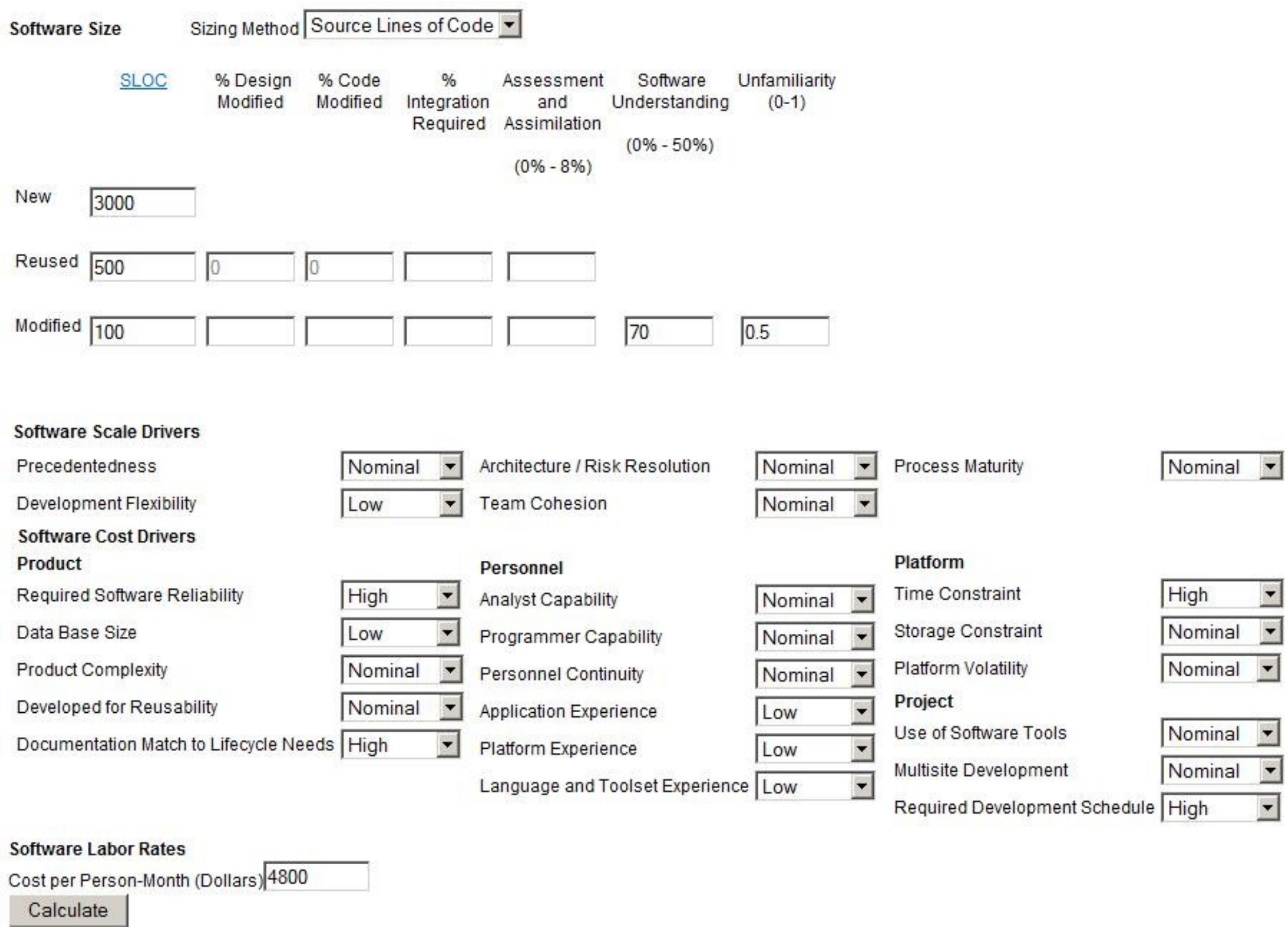

**Figure 15.** Cost estimation with COCOMO II Model, Organic Software mode

**Results**

**Software Engineering**

Effort = 15 Person-months
Schedule = 11 Months
Cost = $76142

**Phase Distribution**

| Phase | Effort (Person-months) | Schedule (Months) | Average Staff | Cost (Dollars) |
|---|---|---|---|---|
| Inception | 1.0 | 1.5 | 0.6 | $4569 |
| Elaboration | 3.8 | 4.5 | 0.9 | $18274 |
| Construction | 12.1 | 7.4 | 1.6 | $57868 |
| Transition | 1.9 | 1.5 | 1.3 | $9137 |

**Staffing by Phase**

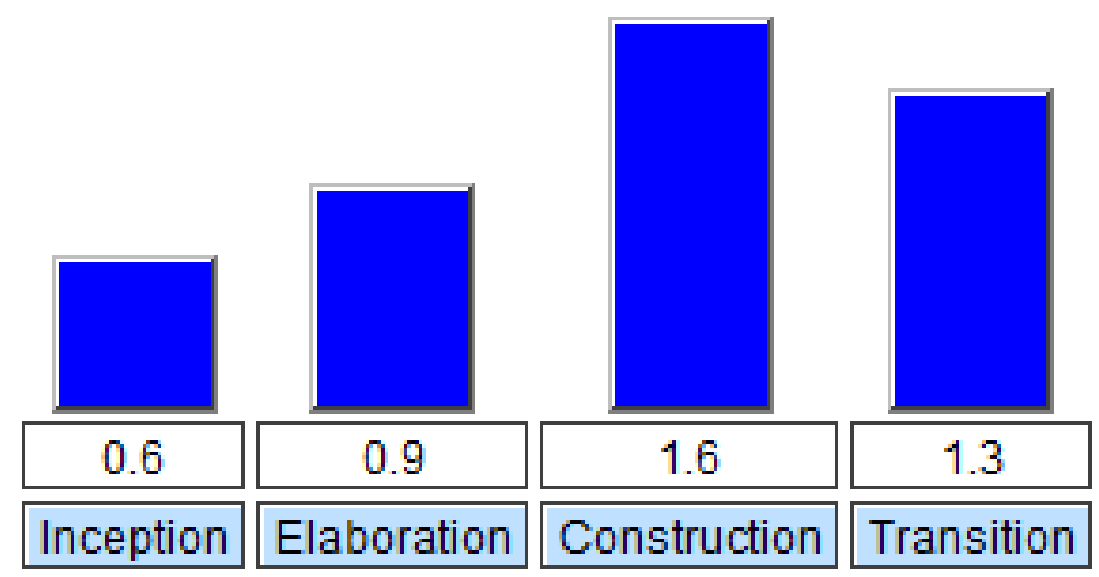

**Software Effort Distribution for RUP/MBASE (Person-Months)**

| Phase/Activity | Inception | Elaboration | Construction | Transition |
|---|---|---|---|---|
| Management | 0.1 | 0.5 | 1.2 | 0.3 |
| Environment/CM | 0.1 | 0.3 | 0.6 | 0.1 |
| Requirements | 0.4 | 0.7 | 1.0 | 0.1 |
| Design | 0.2 | 1.4 | 1.9 | 0.1 |
| Implementation | 0.1 | 0.5 | 4.1 | 0.4 |
| Assessment | 0.1 | 0.4 | 2.9 | 0.5 |
| Deployment | 0.0 | 0.1 | 0.4 | 0.6 |

Your output file is http://csse.usc.edu/tools/data/COCOMO_April_11_2010_10_53_39_48938.txt

Created by Ray Madachy at the Naval Postgraduate School. For more information contact him at rjmadach@nps.edu

**Figure 16.** Results of cost estimation with COCOMO II Model, Organic Software mode

# BIBLIOGRAPHY

Atlee J.M. and Pfleeger S.L. (2010). *Software Engineering: Theory and Practice*. (4th ed.) NJ: Prentice Hall.

COCOMO. In *nasa*. Retrieved Apr. 9, 2010 from http://cost.jsc.nasa.gov/COCOMO.html

COCOMO. In usc. Retrieved Apr. 9, 2010, from http://csse.usc.edu/tools/COCOMOII.php

COCOMO. In *Wikipedia, the free encyclopedia*. Retrieved Apr. 9, 2010, from http://en.wikipedia.org/wiki/COCOMO

IEEE.pdf. In *njit*. Retrieved Mar. 31, 2010 from http://www.cis.njit.edu/~campbell/CIS673/IEEE.pdf

Samples. In *ConcordiaUniversity*. Retrieved Feb. 3 and 13, 2010, from http://users.encs.concordia.ca/~c55414/samples/

Srs.doc. In *itmorelia*. Retrieved Mar. 25, 2010 from http://antares.itmorelia.edu.mx/~jcolivar/documents/srs.doc

Software requirements specs. In *techwr-l*. Retrieved Mar. 25, 2010, from http://www.techwr-l.com/techwhirl/magazine/writing/softwarerequirementspecs.html

UML Use Case Diagrams: tips and FAQ. In *Andrew.* Retrieved April 27, 2010, from http://www.andrew.cmu.edu/course/90-754/umlucdfaq.html#branching

Van Vliet, H. (2008). *Software engineering: Principles and practice* (3rd ed.). Chichester, UK: John Wiley & Sons.